\documentclass[11pt,a4paper]{article}
\usepackage[T1]{fontenc}
\usepackage{flafter}
\usepackage[utf8]{inputenc}
\usepackage{mathptmx,helvet,courier,graphicx,amsmath,amssymb,array}
\newcommand{\toprule}{\hline}
\newcommand{\midrule}{\hline}
\newcommand{\bottomrule}{\hline}
\newcommand{\FloatBarrier}{\clearpage}
\newcommand{\url}[1]{{\small\texttt{\detokenize{#1}}}}
\makeatletter
\long\def\@makecaption#1#2{\vskip 9pt\noindent
  \parbox{\hsize}{\raggedright\small\normalfont\textbf{#1.} #2}\par\vskip 8pt}
\makeatother
\renewcommand{\arraystretch}{1.15}
\title{Clustered Rewiring in Production Networks}
\author{Hiroyasu Inoue$^{1,2}$ \quad Christian Diem$^{3}$ \quad Tobias Reisch$^{3}$\\[0.5em]
\normalsize $^{1}$Graduate School of Information Science, University of Hyogo, Kobe, Japan\\
\normalsize $^{2}$RIKEN Center for Computational Science, Kobe, Japan\\
\normalsize $^{3}$KOF Swiss Economic Institute, ETH Zurich, Zurich, Switzerland\\
\normalsize Correspondence: inoue@gsis.u-hyogo.ac.jp}
\date{}
\begin{document}
\maketitle

\begin{abstract}
\noindent
Production networks change as firms dissolve existing trading relationships and form new ones. We ask how link churn clusters within firms, around churn events, and across connected firms. Using annual Japanese supplier--client records for 2007--2024, our three main comparisons follow a common 2013--2024 sample of 1.45 million firms, with 2.92 million observed dissolutions and 3.84 million formations. We construct references accounting for firms' usual activity, previous link numbers and characteristics, and annual conditions. First, annual counts contain more zeros and more high-count observations than a benchmark fixing each firm's full-period churn total; the count-frequency differences remain significant after Holm correction. Second, mean neighboring counts exceed their benchmarks in all sixteen combinations of direction and dissolution/formation, with Holm-adjusted $p=0.016$. Relative excess is largest on the focal supplier's other outgoing links, reaching 64.5\% for formation around formation. Third, the share of linked pair-years in which both firms exceed their fitted mean churn counts is 27.2\% higher than under a binary model matching firm and year exceedance totals in expectation. This pooled difference has Holm-adjusted $p=0.001$ under a fixed-probability Bernoulli reference. The overall joint excess amount, including zero pair-years, is 64.1\% above an independent-firm count allocation, with Holm-adjusted $p=0.001$. The two references address occurrence and overall amount separately. These results describe clustered rewiring through annual concentration, local co-occurrence, and joint excess activity. They do not identify causal transmission, and annual snapshots do not establish the order of churn events within a year.
\end{abstract}


\section{Introduction}
\suppressfloats[t]
Production networks organize firms' access to inputs and markets, and the productive opportunities available to them. Their evolution matters for productivity and economic growth. Models of endogenous production networks explain how supplier choices and new input combinations shape productivity, and how new production possibilities can support sustained growth \cite{oberfield2018,acemogluazar2020}. Using Belgian firm transactions, Baqaee et al. \cite{baqaee2025} estimate the effects of supplier additions and losses on downstream marginal costs and use those estimates in growth accounting. These studies make the formation and reorganization of relationships economically consequential objects of study.

Capturing changes in industrial structure is important for understanding how the organization of production evolves. Herrendorf et al. \cite{herrendorf2014} review the long-run reallocation of economic activity across agriculture, manufacturing, and services and explain its relevance to development and aggregate productivity. A value-chain perspective also reveals changes across countries and industries: Timmer et al. \cite{timmer2014} trace the value added embodied in manufacturing production and document international fragmentation and shifts in its location and factor composition. Gaggl et al. \cite{gaggl2026} examine long-run changes in US sectoral intermediate-input and investment networks separately. Their measurements and multi-sector analysis show why structural reallocation can have different growth implications across these uses of output. Industrial change thus concerns both the composition of activity and the production connections through which that activity is organized.

Firm-level relationship changes provide a detailed view of one margin of this reorganization. Sectoral shares describe the allocation of activity, while input-output accounts describe aggregate flows between sectors. Within those aggregates, firms can create and dissolve different relationships, and similar aggregate patterns can accompany different sequences of partner changes. Measuring these changes reveals which firms are involved, when they adjust, and whether adjustment is dispersed or concentrated in particular firm-years and connected neighborhoods. This information helps characterize how industrial reorganization unfolds, including changes within industries that sectoral totals do not resolve. We therefore study the temporal and local organization of link churn as a firm-relationship aspect of industrial structural change.

A firm that continues to produce the same products may switch suppliers or clients while maintaining a similar number of trading partners. Introducing new products or discontinuing existing ones may involve changes in several supplier and client relationships. We call the observed disappearance of a trading link dissolution (D), its appearance formation (F), and the two together \emph{churn}. Each D or F is a churn event. We use \emph{rewiring} without requiring a one-to-one match between D and F. Losing five links and adding five others generates substantial churn with no net degree change; adding links without a corresponding loss also contributes. The empirical question is how these changes are distributed through firms' histories and around their positions in the network.

We organize the related literature into four connected strands: relationship turnover and network evolution; search, matching, and expansion; reorganization around disruptions; and models of endogenous network adjustment.

The first strand documents turnover, persistence, firm heterogeneity, and structural dependence in trading relationships. In Japanese firm networks, Fujii et al. \cite{fujii2017} study additions, separations, and continuation counts together with relationship survival, relating them to firm age and growth. Kawakubo and Suzuki \cite{kawakubo2022} document substantial partner renewal among continuing firms and develop a matching model incorporating productivity-related selection. These studies show why a firm that changes many partners in one year cannot automatically be considered unusual: churn varies systematically across firms and stages of their development. Our firm-level question concerns the distribution of a firm's activity across its own years after retaining its full-period total and allowing expected activity to vary with observed annual conditions.

Within this first strand, detailed network-evolution studies connect relationship turnover with the structure of production networks. Atalay et al. \cite{atalay2011} combine US customer microdata with a growth, decay, and attachment model to reproduce features of production-network structure that a purely scale-free account misses. Reisch et al. \cite{reisch2026} use Hungarian monthly transaction data to examine link formation, disappearance, survival, and attachment, and construct a model of network rewiring. Their analysis includes supplier--customer industry-pair turnover patterns. This work establishes much more than the existence of link churn: it characterizes link persistence and the structural regularities of link formation. We build on that evidence by measuring annual count concentration and neighboring excess relative to fitted firm-activity references. A model that reproduces degree distributions, average turnover, or survival can be assessed additionally against these concentration and co-occurrence statistics.

The temporal aggregation and direction of the outcome also matter. Yamano and Sakata \cite{yamano2023} study the metabolism of regional Japanese production networks using changes in firms' partner sets, with industry and temporal descriptions as well as a period-averaged turnover measure. Their undirected dissimilarity measure combines increases, decreases, and replacements. Our analysis instead retains D and F and the supplier--client orientation, asking whether changes accumulate within particular annual intervals and whether different event combinations accompany one another. This distinction also governs the interpretation of industry patterns: industry differences in average partner-set turnover and industry-year differences in firm-level above-benchmark activity concern different units. Neither industry heterogeneity nor a long panel alone is the contribution claimed here.

This first strand also examines how network structure helps explain relationship changes. Matous and Todo \cite{matoustodo2016} use stochastic actor-oriented models to relate link formation and maintenance among Japan's largest firms around the 2011 earthquake to local interdependencies and geography. Their subsequent automotive study jointly examines network evolution and organizational performance \cite{matoustodo2017}. Arata and Mundt \cite{arata2019} model partner changes in a larger Japanese network using local topology and firm attributes. Krichene et al. \cite{krichene2019} estimate an exponential random graph model of adjacency within identified production-network communities. Asai and Nirei \cite{asainirei2025} estimate a dyadic link-formation model using a conditional logit over quadruples of firms. They use the estimates to study the implications of common partners in general equilibrium. These studies provide precedents for structural dependence in trading relationships. Our outcome is the amount and type of other D/F around a realized focal churn event, and the joint above-benchmark activity of connected firms. We do not treat independently allocated firm activity as a null preserving every structural mechanism of partner selection.

The second strand explains how search, matching, and new opportunities shape relationship adjustment. Search and matching research shows why replacement and expansion can be gradual and geographically uneven. Following unanticipated supplier bankruptcies, Miyauchi \cite{miyauchi2024} finds that Japanese firms rematch with alternative suppliers progressively, and faster where alternative suppliers are geographically denser. Monarch \cite{monarch2022} combines US--China trading relationships with a dynamic supplier-choice model to estimate switching frictions. Bernard et al. \cite{bernard2019} use improved passenger access following a Japanese high-speed-rail opening to study new buyer--seller relationships and firm performance. These studies establish formation and rematching as outcomes shaped by search conditions. The initiating bankruptcy in the first study differs from our changes between firms recorded in both consecutive years; subsequent rematching can nevertheless involve such continuing firms. Our annual concentration statistics do not by themselves estimate search or switching costs.

This second strand also examines relationship formation associated with positive opportunities. Randomized business meetings studied by Cai and Szeidl \cite{cai2018} increase partnerships and business performance among young Chinese firms; these partnerships include suppliers and clients but are broader than intermediate-input links. Demir et al. \cite{demir2024} find that Turkish firms exposed to stronger demand from richer export destinations form trading relationships with higher-wage suppliers and clients and shift spending toward higher-wage continuing suppliers. Furusawa et al. \cite{furusawa2018} relate global sourcing by Japanese firms to domestic supplier retention and additions, with responses differing across firms and inputs. Carvalho and Voigtl\"ander \cite{carvalhovoigtlander2014} study input diffusion through existing supplier connections. Together, these studies connect evolving relationships with access to new markets, upgrading, and input adoption. Our joint consideration of D and F accommodates these expanding activities alongside contraction and substitution.

Evidence connecting renewal with growth also goes beyond the choice of a new partner. Hisano et al. \cite{hisano2017} use a decade of Japanese buyer--seller networks and a structural-equation framework to examine sales growth and the role of link renewal in the propagation of positive and negative influences. Their counterfactual comparisons change the network through which estimated shocks travel. The economic response being evaluated is growth, whereas our dependent objects are the timing, direction, and co-occurrence of recorded link churn itself. This difference in outcome is consequential: growth comovement at nearby firms need not imply that their relationship changes cluster, and clustered churn does not by itself establish a growth benefit or loss.

The third strand examines reorganization around adverse events and disruptions, when several connected firms may adjust. Work on aggregate fluctuations, disasters, and firm-level shock propagation establishes the economic consequences of supplier--client interdependence \cite{carvalho2014,acemoglu2012,barrot2016,inoue2019,carvalho2021}. More directly concerning relationship changes, P\'alovics et al. \cite{palovics2021} compare firms exposed to partners' sharp sales declines with matched firms in Slovenian payment networks and find subsequent adjustments in trading partners. Korovkin et al. \cite{korovkin2026} use Ukrainian shipment data around the 2014 conflict. For partners outside the conflict areas, firms with high supplier exposure show a relative increase in supplier-link counts and tend to reduce buyer-link counts, whereas firms with high buyer exposure show relative decreases in both counts. These findings document network-local reorganization and motivate separating the direction and type of accompanying changes. We use the observed D and F throughout the annual panel as focal events. The comparisons describe the changes that accompany those events across the range of conditions in the panel.

Within this third strand, the pandemic literature is especially close in data and outcomes. Saito et al. \cite{saitotong2026} use TSR records for 2015--2023 to examine supplier and buyer additions and deletions with firm effects and controls for employment and existing degree. Their event-study estimates describe within-firm changes relative to the pre-pandemic period. Thus neither D/F counts nor adjustment for persistent firm differences is new here. Our comparisons instead examine the entire annual count distribution relative to a fixed-total allocation, the four-direction neighboring counts conditional on realized focal D/F, and joint excess at background neighbors. Khanna et al. \cite{khanna} use transaction records from one Indian state and differences in exposure to lockdowns affecting suppliers to study separation, production outcomes, and subsequent supplier selection. Their evidence of later additions of larger, better-connected suppliers shows how a disruption can be followed by selective formation; it does not supply the benchmark for our pooled concentration statistics.

Other adjustment studies show why recorded link churn must be distinguished from changes in sourcing volumes or geography. Handley et al. \cite{handley2024} distinguish the discontinuation of buyer--supplier relationships, reduced formation of new ones, and firm exits from import markets in the 2018--2019 U.S. trade war. Niu et al. \cite{niu2025} document supplier additions and reallocation in US maritime imports during the pandemic, with many new suppliers located in existing sourcing countries. Cajal-Grossi et al. \cite{cajal2023} distinguish supplier retention, sourcing-country exit, and purchase values for apparel buyers with different sourcing strategies. Son et al. \cite{son2021} compare changes in affected Japanese firms' network degrees and ego-network density after the 2011 earthquake. Survey-based substitution in transaction volumes, such as that examined by Todo et al. \cite{todo2023}, does not by itself establish the creation of a new dyad. These outcomes describe related dimensions of reorganization, but cannot all be interpreted as counts of binary link births and deaths or as evidence of a temporal chain of changes.

The fourth strand develops models of endogenous network adjustment, providing mechanisms against which these empirical regularities can be evaluated. Huneeus \cite{huneeus2018} studies dynamic network adjustment with frictions using Chilean firm transactions. Dhyne et al. \cite{dhyne2023} introduce fixed link-formation costs and calibrate their model to Belgian transactions. In their trade-shock simulations, domestic link turnover is substantial and welfare changes differ only moderately from the fixed-link case. Arkolakis et al. \cite{arkolakis2023} model supplier and buyer search across locations and evaluate trade and welfare with endogenous matching. These models link changes in who trades with whom to aggregate outcomes. Our reference comparisons measure how concentrated the observed link changes are after accounting for specified patterns of firm activity.

Within this fourth strand, the meaning of a cascade also differs across models. Endogenous relationship dissolution can reinforce contraction in Acemoglu and Tahbaz-Salehi \cite{acemoglutahbaz2025}, while supplier replacement and adaptation can amplify fluctuations in K\"onig et al. \cite{konig2022}. Baqaee \cite{baqaee2018} and Taschereau-Dumouchel \cite{taschereau2026} study failures or complementary operating decisions, including margins involving entry, exit, or shutdown. Elliott et al. \cite{elliott2022} connect endogenous sourcing decisions with fragility. These are theoretical reasons to study interdependent adjustment, rather than empirical demonstrations that the annual D/F in our data form causal cascades. We count changes between firms recorded as active and with retained trading relationships in both adjacent snapshots. This allows links involving continuing firms to change in response to disruptions elsewhere. A large connected set of links undergoing churn alone cannot identify any of these mechanisms.

These four strands motivate a common question about how strongly link churn clusters in time and around firms after accounting for firms' usual activity and annual conditions. Existing work measures relationship turnover, formation, persistence and partner choice, or examines adjustment to particular shocks and new opportunities. We therefore provide an empirical characterization of three forms of clustering: concentration of a firm's churn into particular years, neighboring D/F around churn events, and joint above-benchmark churn at connected firms. Each comparison uses an explicit reference specifying which features of firm activity it preserves \cite{gauvin2022}.

At the firm-year level, Analysis~1 fixes each firm's full-period churn total and allocates it across eligible years using probabilities derived from a fitted count model. It therefore asks about the annual distribution given that firm's overall activity. Analysis~2 starts from a realized D or F and measures other D/F on the supplier's and client's incoming and outgoing links separately. It compares mean neighboring counts with the levels predicted from firms' usual activity. In Analysis~3, a binary model for above-mean churn matches each firm's and each year's exceedance totals in expectation. The count allocation supplies the reference for the zero-inclusive overall joint excess amount. These different observation units and preserved quantities make the comparisons complementary, not an additive decomposition under one common null. We also describe how firm-level churn varies across industries and years and compare joint exceedance across industry pairs.

The findings show more zero and high-count firm-years than predicted, excess mean neighboring counts across all directions and D/F combinations, and both more joint exceedances and a larger overall co-excess amount at transaction neighbors relative to their respective references. Both pooled differences are significant under their respective fixed-fit simulation references. We refer to this empirical organization as \emph{clustered rewiring}. The additional temporal comparisons examine recurrence and next-year continuation. These patterns provide empirical requirements beyond reproducing mean turnover or the static network: the allocation of changes through time, their D/F composition around churn events, and their joint occurrence and amount at connected firms.

\section{Data and measurement}
\suppressfloats[t]
\subsection{Annual transaction networks and observed changes}
Tokyo Shoko Research (TSR) collects information on firms' suppliers and clients through questionnaires answered by firms \cite{krichene2019}. We use its company and trading-partner records for 2007--2024. Appendix~\ref{app:annual-network} reports annual firm and link counts for this full period (Figure~\ref{fig:annual-network} and Table~\ref{tab:annual-network}) and firm-level churn statistics for the main 2013--2024 panel (Table~\ref{tab:firm-churn-statistics}). A directed link points from a supplier to a client. We include a relationship when either firm reports it, count repeated reports of the same directed pair once, and exclude self-links. Combining both sides' reports can give a firm more links than appear on its own list.

The earlier TSR reporting format requested up to 24 major suppliers and 24 major clients \cite{goto2020}. In the files used here, a firm's own lists can exceed 24 from 2021 onward. Reporting continues to focus on major trading partners. These restrictions concern reported relationships rather than an exhaustive census of every transaction.

Let $E_t$ be the set of observed directed links in year $t$. We measure link churn between two consecutive annual snapshots:
\begin{equation}
D_t=E_{t-1}\setminus E_t,\qquad F_t=E_t\setminus E_{t-1}.
\label{eq:events}
\end{equation}
\begin{samepage}
The events in Equation~\eqref{eq:events} are restricted to links whose two endpoint firms have company records and are active in both $t-1$ and $t$ \cite{tsrstatus}. Each endpoint must also have at least one retained supplier or client relationship in each snapshot. The relationship need not be with the same partner in both years. We assess this condition before selecting the transition's endpoint cohort, then count D/F links between qualifying firms. Eligibility is assessed separately for each annual transition; firms need not be observed throughout the entire study period.
\par
\end{samepage}

Throughout, \emph{same-year} means the same transition between two annual snapshots, not exact simultaneity.

\subsection{Samples and observation units}
Analysis~1 (Temporal concentration of firm-level churn; Section~\ref{sec:temporal-concentration}) examines how churn is distributed across years within each firm. Analysis~2 (Neighboring churn; Section~\ref{sec:neighboring-churn}) examines what other D/F changes occur around a churn event in the same year. Analysis~3 (Joint excess churn at transaction neighbors; Section~\ref{sec:joint-excess}) examines whether transaction neighbors jointly exceed their own expected churn more often and with a greater overall amount than firm-level benchmarks predict. These three comparisons use firm-years, focal link-years, and background firm-pair-years, respectively.

Analyses~1--3 use the same eligible 2013--2024 panel: 12,194,211 firm-years from 1,450,395 firms, including zeros (Table~\ref{tab:samples}). The extensions in Appendices~\ref{app:firm-multiyear}--\ref{app:joint-multiyear} retain this annual eligibility rule. Firm-level and directional regressions cover 2008--2024; fixed-total firm overlaps and industry-year descriptions follow a common cohort eligible in all seventeen transitions. Link-level controlled comparisons cover 2013--2024. Joint excess requires eligibility in both compared annual intervals within that period.

We retain the earlier snapshots for the underlying history and for the 2008--2024 comparison of firm activity by industry and year in Section~\ref{sec:industry-patterns} and churn participation in Appendix~\ref{app:participation}.

At a firm $v$, gross churn in year $t$ combines the four incoming/outgoing D/F counts:
\begin{equation}
C_{vt}=D^{\mathrm{out}}_{vt}+D^{\mathrm{in}}_{vt}
       +F^{\mathrm{out}}_{vt}+F^{\mathrm{in}}_{vt}.
\label{eq:gross}
\end{equation}
Across the panel, outgoing and incoming D counts each sum to 2,917,266, while outgoing and incoming F counts each sum to 3,844,702. The 6,761,968 unique D/F churn events therefore produce 13,523,936 endpoint incidences: each event contributes one count to each endpoint in Equation~\eqref{eq:gross}. The firm-level comparison uses one gross count per firm-year, retaining zeros.

To examine joint excess, we consider pairs of firms connected by a transaction in either the previous or current year. This includes continuing relationships as well as D and F. We count each pair once per year, regardless of transaction direction. The occurrence and amount comparisons use the resulting 59,099,713 pair-years.

\begin{figure}[tbp]
\centering
\includegraphics[width=\linewidth,height=0.80\textheight,keepaspectratio]{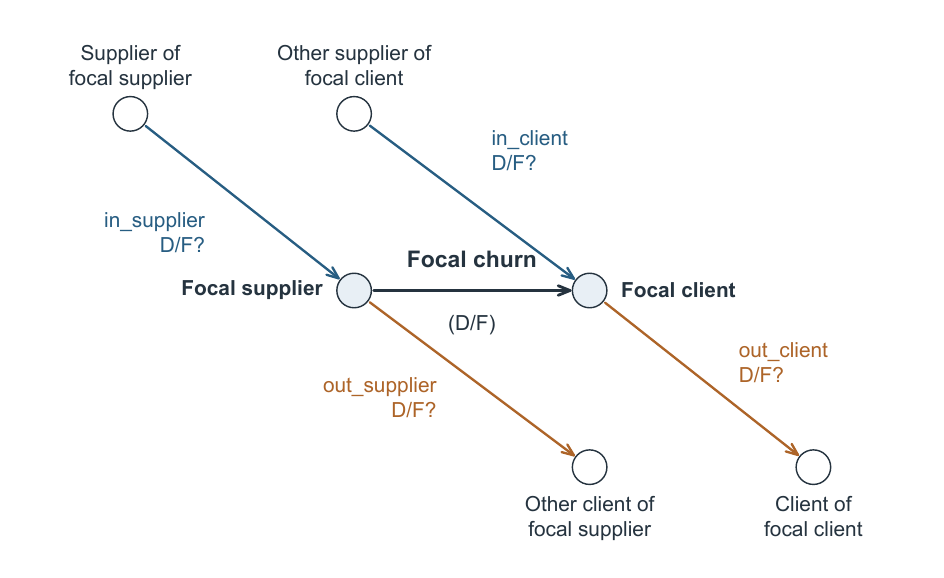}
\caption{Four neighboring directions around a focal supplier--client link. Arrows show the direction of trade. The focal link undergoes D or F; D/F? marks the other links checked for churn, which need not occur. The focal churn event itself is excluded from neighboring counts.}
\label{fig:directions}
\end{figure}

For neighboring churn, we select a focal supplier--client link $i\rightarrow j$ undergoing D or F. There are 2,917,266 focal D link-years and 3,844,702 focal F link-years. A neighboring link can enter or leave either endpoint. Figure~\ref{fig:directions} distinguishes four positions: the focal supplier's own suppliers (\texttt{in\_supplier}), its other clients (\texttt{out\_supplier}), the focal client's other suppliers (\texttt{in\_client}), and its own clients (\texttt{out\_client}). Four positions, two focal types, and two neighboring types give sixteen comparisons. Each focal type's eight comparisons use the same link-years, including those with zero neighboring counts. The focal churn event is excluded from matching counts. Churn on the reverse link $j\rightarrow i$ can enter both \texttt{in\_supplier} and \texttt{out\_client}; directions are compared separately rather than summed.

\begin{table}[tbp]
\centering\small
\caption{Observation units and sample sizes in the three main comparisons, 2013--2024. All denominators include observations with zero counts or zero joint excess.}
\label{tab:samples}
\begin{tabular}{@{}>{\raggedright\arraybackslash}p{4.2cm}>{\raggedright\arraybackslash}p{3.3cm}>{\raggedleft\arraybackslash}p{2.2cm}>{\raggedright\arraybackslash}p{5.0cm}@{}}
\toprule
Analysis & Observation unit & Number & Quantity examined \\
\midrule
Analysis~1: Temporal concentration of firm-level churn & Firm-year & 12,194,211 & Distribution of annual churn counts within firms \\[4pt]
Analysis~2: Neighboring churn & Focal D link-year & 2,917,266 & Neighboring D/F counts in four directions around dissolved links \\[4pt]
Analysis~2: Neighboring churn & Focal F link-year & 3,844,702 & Same comparison around newly formed links \\[4pt]
Analysis~3: Joint excess churn at transaction neighbors & Background firm-pair-year & 59,099,713 & Frequency and overall amount of joint firm-specific excess \\
\bottomrule
\end{tabular}
\end{table}

\FloatBarrier
\section{Analyses and results}
\suppressfloats[t]
\subsection{Analysis~1: Temporal concentration of firm-level churn}
\label{sec:temporal-concentration}
\subsubsection{Firm-level count model and allocation benchmark}
We ask whether a firm's churn is more concentrated in particular years than expected after accounting for its usual activity, its characteristics, and overall annual activity. Accounting for these factors is necessary because firms with more links or persistently high churn can record many changes even without an unusually active year. Moreover, spreading their changes evenly across years would still ignore variation in degree and annual conditions.

We fit a Poisson regression to the observed gross counts in Equation~\eqref{eq:gross} \cite{mccullagh1989}. For firm $v$ in year $t$, let $Y_{vt}$ denote the count as a random variable, with observed value $C_{vt}$. The model is
\begin{equation}
Y_{vt}\sim\operatorname{Poisson}(\mu_{vt}),\qquad
\log\mu_{vt}=a_v+b_t+\boldsymbol\theta^{\mathsf T}\mathbf Z_{v,t-1}.
\label{eq:poisson}
\end{equation}
The firm effect $a_v$ accounts for persistent differences in usual churn, and the year effect $b_t$ for overall annual activity. The controls $\mathbf Z_{v,t-1}$ are previous total incoming-plus-outgoing degree, employment, and capital. They enter as $\log(1+x)$, with indicators for missing employment and capital. Appendix~\ref{app:regressions} reports the coefficients. Firms with no churn throughout their eligible years do not enter coefficient estimation because they have no finite Poisson firm effect; they remain in the comparisons with zero counts.

We use the fitted counts to allocate each firm's observed total across its eligible years. Let $\mathcal T_v$ be those years and define
\begin{equation}
N_v=\sum_{s\in\mathcal T_v}C_{vs},\qquad
p_{vt}=\frac{\widehat\mu_{vt}}{\sum_{s\in\mathcal T_v}\widehat\mu_{vs}}
\quad (N_v>0).
\label{eq:allocation-probability}
\end{equation}
Firm effects make the fitted sum equal to $N_v$ within numerical accuracy. Normalization ensures that the probabilities sum to one. Each count falls in year $t$ with probability $p_{vt}$, giving the annual reference
\begin{equation}
\widetilde C_{vt}\mid N_v\sim\operatorname{Binomial}(N_v,p_{vt}),
\qquad t\in\mathcal T_v.
\label{eq:annual-binomial}
\end{equation}
Firms with $N_v=0$ retain zero counts and means throughout; they need no annual probabilities. The same convention applies to zero totals in the directional models below. Aggregate annual totals match in expectation.

Because the observed and reference totals and the number of firm-years agree, their overall means must agree. The question is how annual counts are distributed. We calculate the probability of each churn count for every firm-year. Averaging these probabilities over all 12,194,211 firm-years, including those with zero observed churn, gives the expected distribution shown in Figure~\ref{fig:firm-distribution}.

To assess departures from the fitted count distribution, we generate 1,999 datasets from the allocation benchmark. In each dataset, we independently assign each of a firm's $N_v$ counts to one eligible year using the fitted probabilities $p_{vt}$. This keeps each firm's full-period total fixed and makes its annual counts dependent across years. The annual allocation probabilities fitted to the observed data are held fixed throughout.

For each churn count or count range, we compare the observed fraction of firm-years with the fractions from the 1,999 generated datasets. For the upper and lower tails, we count simulated fractions at or above and at or below the observed fraction, respectively. Each one-sided Monte Carlo p-value is one plus the corresponding tail count, divided by 2,000. The two-sided p-value doubles the smaller tail value and is capped at one. Its smallest attainable value is 0.001. We apply Holm correction across 21 distinct comparisons covering the ranges in Figure~\ref{fig:firm-distribution}(a), individual counts with a pooled upper tail, and the alternative groupings in Appendix~\ref{app:reference}. That appendix reports all 21 results and a supplementary whole-distribution diagnostic.

\subsubsection{Results}
\begin{figure}[tbp]
\centering
\includegraphics[width=0.85\linewidth,height=0.80\textheight,keepaspectratio]{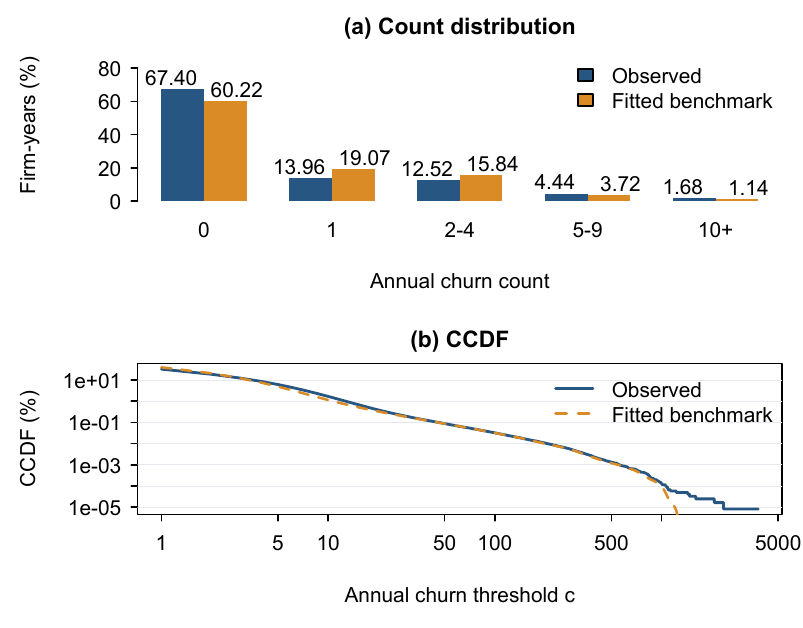}
\caption{Annual gross churn at firms, 2013--2024. (a) Percentages in count ranges. (b) Fractions with at least the indicated count, on logarithmic axes. The denominator in both panels is all 12,194,211 firm-years, including those with zero churn. The benchmark is the analytic binomial mixture implied by Equation~\eqref{eq:annual-binomial}. Each churn event enters both endpoint firms' gross counts. Count-range tests for panel (a) are reported in Tables~\ref{tab:firm-count-frequencies} and~\ref{tab:firm-count-sensitivity}, with Holm correction across all 21 comparisons.}
\label{fig:firm-distribution}
\end{figure}

The observations contain more zero-churn firm-years and more high-count firm-years than predicted (Figure~\ref{fig:firm-distribution}). Zero counts occur in 67.40\% of observations, compared with 60.22\% expected. Ten or more changes occur in 1.684\%, compared with 1.141\% expected, a ratio of 1.476. Counts of one to four are less frequent than predicted. With the same overall mean of 1.109 changes per firm-year, churn is present in fewer firm-years but more concentrated in high-count observations.

All five ranges in Figure~\ref{fig:firm-distribution}(a) differ significantly from their references. Each raw two-sided Monte Carlo p-value reaches the minimum attainable value, $p=0.001$, and each Holm-adjusted p-value is 0.021. The correction covers the same family of 21 comparisons, including the finer and alternative ranges. The finer comparisons show deficits at each count from 1 through 4, and excesses at each count from 5 through 9 and in the 10+ tail. Pooling changes the size of the differences, but using 2--3 and 4--9, or 2--5 and 6--9, still gives a lower-range deficit and an upper-range excess. The 8+ and 12+ tails are also above expectation (Tables~\ref{tab:firm-count-frequencies} and~\ref{tab:firm-count-sensitivity}). The supplementary whole-distribution diagnostic is reported in Appendix~\ref{app:reference}.

The reference allocates counts independently across firms, whereas every observed churn event contributes to two endpoint counts. The result therefore establishes departure from this firm-level allocation model, not from a reference preserving all dependencies of a complete transaction network. Section~\ref{sec:industry-patterns} shows how the count concentration varies across industries.

Appendix~\ref{app:firm-multiyear} examines continuation beyond one annual interval. Observed next-year gross-churn recurrence is 59.45\%, compared with 47.26\% under the degree-and-year benchmark and 60.25\% after adding firm effects. Within each measure's group with at least ten first-year changes, next-year mean counts exceed their firm-effect predictions by 0.215 D, 0.767 F, and 0.214 gross churn events per firm-year. A separate fixed-total comparison in the new common linked cohort gives fewer pairs of years with churn at the same firm than expected, at separations of one to five years. These results distinguish pooled recurrence, continuation after large bursts, and the allocation of activity across a firm's history; the regression-based gaps are descriptive in-sample comparisons.

\FloatBarrier
\subsection{Analysis~2: Neighboring churn}
\label{sec:neighboring-churn}
\subsubsection{Focal churn events, neighboring counts, and their benchmark}
Analysis~1 (temporal concentration; Section~\ref{sec:temporal-concentration}) examines how the distribution of gross churn counts per firm-year differs from its benchmark, without distinguishing link directions. Here we ask whether D or F on a focal link is accompanied by more neighboring D or F in the same year than the benchmark predicts, and how this association varies across directions. For each observed focal churn event, we record the numbers of neighboring D and F separately in each of the four directions in Figure~\ref{fig:directions}. We then sum the neighboring counts and divide by the number of focal churn events, separately for focal D and focal F.

We first fit the specification in Equation~\eqref{eq:poisson} separately for four count types, indexed by $r$: outgoing D, incoming D, outgoing F, and incoming F. Each fit uses matching previous outgoing or incoming degree; the other controls and firm and year effects are unchanged. The four fits use the same main panel of 12,194,211 firm-years from 1,450,395 firms and have their own coefficients (Appendix~\ref{app:regressions}). For each count type, firms with no churn of that type throughout their eligible years do not enter coefficient estimation because they have no finite Poisson firm effect. They remain in the comparisons with zero counts and means. For each type, we construct its firm-specific full-period total $N_v^r$ and annual probabilities $p_{vt}^r$ as in Equation~\eqref{eq:allocation-probability}.

A firm's count may include the focal churn event itself; we exclude it when counting churn events on neighboring links. For a focal link $\ell=(i\rightarrow j)$ in year $t$, the chosen direction $d$ and neighboring D/F mark determine the incoming or outgoing count type $r$. Let $v(d)=i$ for the two supplier directions and $v(d)=j$ for the two client directions. $C^{r}_{v(d)t}$ counts type-$r$ churn events at firm $v(d)$ in year $t$. We set $\delta_{\ell}^{r,d}=1$ if this count includes the focal churn event and $0$ otherwise. This exclusion is one only when the neighboring and focal D/F marks match and $d$ is \texttt{out\_supplier} or \texttt{in\_client}; otherwise it is zero. The observed neighboring count $K_{\ell t}^{r,d}$ and its benchmark mean $m_{\ell t}^{r,d}$ are
\begin{equation}
K_{\ell t}^{r,d}=C^{r}_{v(d)t}-\delta_{\ell}^{r,d},\qquad
m_{\ell t}^{r,d}=(N_{v(d)}^r-\delta_{\ell}^{r,d})p_{v(d)t}^r.
\label{eq:neighbor-count}
\end{equation}
The full reference distribution underlying Equation~\eqref{eq:neighbor-count} is binomial with parameters $N_{v(d)}^r-\delta_{\ell}^{r,d}$ and $p_{v(d)t}^r$. Thus excluding the focal churn event removes one from the matching full-period total as well as from the observation; it does not subtract one from the fitted annual mean.

We average the observed counts and their benchmark means over the same set of focal events. Four directions crossed with two focal and two neighboring types give sixteen comparisons. In Analysis~1 (temporal concentration; Section~\ref{sec:temporal-concentration}), fixed firm totals make the overall observed and expected means equal. Here we average per focal churn event, so the observed and expected neighboring means can differ even with fixed firm totals.

To test the mean differences, we use 1,999 allocations with the fitted parameters fixed. For each firm and count type, we keep the full-period total fixed and assign each count to one eligible year using the fitted annual probabilities. Counts are allocated independently across firms and count types. In each allocation, we calculate the simulated and benchmark neighboring means using the focal-event counts generated in that allocation. Whenever a comparison uses a count of a given type for the same firm and year, it uses the same simulated count.

For each comparison, we calculate upper- and lower-tail Monte Carlo p-values by adding one to the corresponding exceedance count and dividing by 2,000. The two-sided value is twice the smaller tail value, capped at one. Holm correction controls the family of sixteen comparisons. A separate global statistic is the maximum absolute percentage mean difference across them. The tests condition on the fitted probabilities and do not refit the regressions.

\subsubsection{Results}
All sixteen mean neighboring counts exceed their benchmarks (Appendix Table~\ref{tab:neighbors}). Each two-sided Monte Carlo $p$ is 0.001, the minimum attainable value, and each Holm-adjusted $p$ is 0.016. The global maximum test gives the minimum attainable upper-tail Monte Carlo $p=0.0005$. These tests compare observed and expected mean neighboring counts.

\begin{figure}[tbp]
\centering
\includegraphics[width=0.85\linewidth,height=0.80\textheight,keepaspectratio]{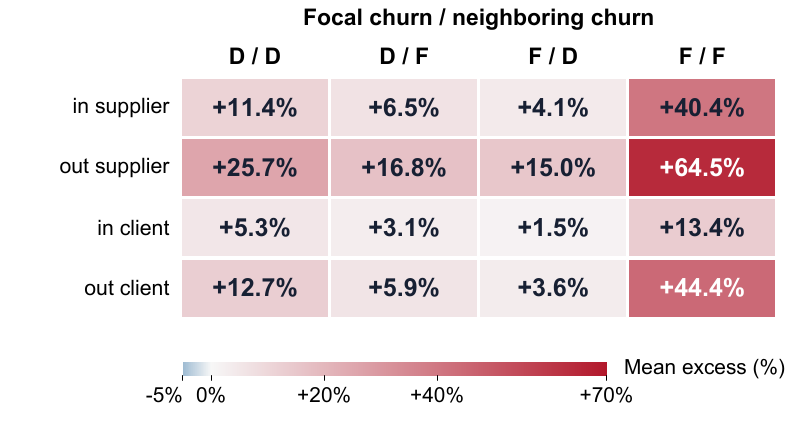}
\caption{Percentage excess of observed mean neighboring counts over expected means in Appendix Table~\ref{tab:neighbors}, computed from unrounded values. The common scale spans $-5$\% to $+70$\%; color represents relative mean excess, not statistical significance. At each endpoint of the focal link, relative excess is larger on outgoing links than on incoming links in all four D/F combinations. Directions refer to Figure~\ref{fig:directions}.}
\label{fig:neighbor-excess}
\end{figure}

Relative excess is systematically larger for same-type D/D and F/F combinations than for mixed D/F and F/D combinations within each direction; F/F is the largest. For focal F with neighboring F on the focal supplier's other outgoing links (\texttt{out\_supplier}), the observed mean is 13.010 against 7.910 expected: 5.100 additional formations per focal F, or 64.5\%. For focal D with neighboring D in this direction, the means are 4.737 and 3.767, an excess of 25.7\%. Across directions, the focal supplier's other outgoing links (\texttt{out\_supplier}) have the largest relative excess in every combination (Figure~\ref{fig:neighbor-excess}). At each endpoint of the focal link, relative excess is larger on outgoing links than on incoming links in all four D/F combinations.

Incoming counts at the focal client can be large in absolute terms while having smaller relative excess. Large raw counts and large departures from the benchmark therefore describe different patterns. The complementary cumulative distributions in Appendix~\ref{app:neighbor-distributions} also show that the mean excess is not an excess at every count threshold: some curves cross.

These comparisons show local same-year co-occurrence relative to a benchmark that accounts for firm characteristics and firm and year effects and fixes each firm's full-period counts. The four directions distinguish upstream and downstream relationships. Annual snapshots do not reveal which change occurred first within the same annual interval; changes can occur together during reorganization without one causing the others. The results are averages per focal churn event, so firms with more focal events contribute more observations.

Table~\ref{tab:neighbor-next-allocation} in Appendix~\ref{app:multiyear-neighboring} reports neighboring churn in the year after a focal churn event. Under the same fixed-total benchmark, neighboring D after focal D and neighboring F after focal F exceed their expected means in all four directions. For focal F, outgoing F at the focal supplier is 24.5\% above its reference. By contrast, neighboring F after focal D and neighboring D after focal F are below expectation in all four directions. Separate link-level regressions in Appendix~\ref{app:neighbor-distributed-lag} compare lagged neighboring churn with and without same-year churn. Their coefficient and lag-sum patterns describe associations; they do not identify causal transmission.

\FloatBarrier
\subsection{Analysis~3: Joint excess churn at transaction neighbors}
\label{sec:joint-excess}
\subsubsection{Background pairs and firm-specific excess}
Analysis~1 (temporal concentration; Section~\ref{sec:temporal-concentration}) examines concentration within firm histories, while Analysis~2 (neighboring churn; Section~\ref{sec:neighboring-churn}) starts from a churn event. We now ask two questions about transaction neighbors: do they exceed their own expected counts together more often than their individual exceedance frequencies predict, and is their overall joint excess amount larger than a count allocation predicts? These questions require different references. For occurrence, we match each firm's number of years with excess churn and each year's number of firms with excess churn, in expectation. For amount, we fix each firm's full-period churn total and allocate those counts across its eligible years.

For each year $t$, the background $\mathcal P_t$ consists of unordered pairs of eligible firms linked by a transaction in $t-1$ or $t$, retaining continuing relationships, D, and F. Each pair is counted once per year, regardless of transaction direction. Across years, this gives $M=\sum_t|\mathcal P_t|=59{,}099{,}713$ pair-years. We retain firms with zero observed churn in the comparison.

We use the annual allocation probabilities $p_{vt}$ from the gross-count model in Analysis~1 (Equation~\eqref{eq:allocation-probability}) to obtain the firm-specific mean $m_{vt}=N_vp_{vt}$. We first mark whether a firm's observed count exceeds this mean:
\begin{equation}
Z_{vt}=\mathbf 1\{C_{vt}>m_{vt}\}.
\label{eq:exceedance-indicator}
\end{equation}
Its positive excess and the amount jointly present at a pair are
\begin{equation}
e_{vt}=\max\{C_{vt}-m_{vt},0\},\qquad
w_{uvt}=\min\{e_{ut},e_{vt}\}.
\label{eq:excess}
\end{equation}
A large excess at one endpoint cannot compensate for none at the other. Here, excess means any amount above the fitted mean; it does not imply statistical significance for an individual firm-year. An observed change in the transaction relationship between two firms contributes to both firms' counts.

\subsubsection{Binary occurrence reference}
We fit a binary logit model to the indicator in Equation~\eqref{eq:exceedance-indicator}. Let $q_{vt}$ be its fitted probability:
\begin{equation}
\log\frac{q_{vt}}{1-q_{vt}}
=\alpha_v+\gamma_t+\boldsymbol\beta^{\mathsf T}\mathbf x_{v,t-1}.
\label{eq:exceedance-logit}
\end{equation}
The control vector $\mathbf x_{v,t-1}$ contains the same five varying covariates used in the gross-count fit: previous total incoming-plus-outgoing degree, employment, and capital, each as $\log(1+x)$, and employment and capital missingness indicators. Firm and year fixed effects account for persistent exceedance frequency and aggregate annual conditions. Here, $q_{vt}$ is the fitted probability that a firm's churn count exceeds its expected count.

The finite fit uses 937,299 firms, comprising 9,302,579 firm-years. The 513,096 firms with no exceedance in any eligible year, comprising 2,891,632 firm-years, have the boundary prediction $q_{vt}=0$ and remain in every relevant background denominator. Appendix~\ref{app:regressions} reports the coefficients and fit details.

\begin{samepage}
For firm $v$, let $\mathcal T_v$ be its eligible years and $K_v=\sum_{t\in\mathcal T_v}Z_{vt}$ its observed number of years with excess churn. The fitted model matches the observed totals in expectation for each firm and each year.
\begin{equation}
\sum_{t\in\mathcal T_v}q_{vt}=K_v,\qquad
\sum_{v:t\in\mathcal T_v}q_{vt}=\sum_{v:t\in\mathcal T_v}Z_{vt}.
\label{eq:exceedance-margins}
\end{equation}
\end{samepage}
The overall observed and expected totals are both 2,786,020 exceedance firm-years. We next compare how often linked firms exceed together after matching these firm and year totals in expectation.

For a pair $\{u,v\}$ in year $t$, $Z_{ut}Z_{vt}$ equals one if both firms exceed their fitted means and zero otherwise. Under the binary reference, the two firms' indicators are independent, so the probability that both exceed is $q_{ut}q_{vt}$. Summing over the yearly sets $\mathcal P_t$ gives the observed and expected joint-exceedance counts.
\begin{equation}
A_{\mathrm{obs}}=\sum_t\sum_{\{u,v\}\in\mathcal P_t}Z_{ut}Z_{vt},\qquad
A_{\mathrm{bin}}=\sum_t\sum_{\{u,v\}\in\mathcal P_t}q_{ut}q_{vt}.
\label{eq:coexceedance-count}
\end{equation}

To test whether joint exceedance is more frequent than predicted, we run 1,999 simulations on the same background. In each repetition $b$, we draw one exceedance indicator $\widetilde Z_{vt}^{(b)}\sim\operatorname{Bernoulli}(q_{vt})$ for every eligible firm-year, independently across firms and years. Pairs with a firm in common can have dependent outcomes because they use the same simulated indicator for that firm in that year. The simulated totals vary across draws around the firm and year totals in Equation~\eqref{eq:exceedance-margins}.

Let $A^{(b)}$ be the simulated joint-exceedance count in repetition $b$. The upper-tail Monte Carlo probability is $[1+\sum_{b=1}^{1999}\mathbf 1\{A^{(b)}\geq A_{\mathrm{obs}}\}]/2000$.

\subsubsection{Count-allocation reference for overall amount}
We measure the overall amount by averaging the joint positive excess $w_{uvt}$ from Equation~\eqref{eq:excess} across all background pair-years, including those with zero joint excess.
\begin{equation}
T=\frac{1}{M}\sum_t\sum_{\{u,v\}\in\mathcal P_t}w_{uvt}.
\label{eq:coexcess-statistic}
\end{equation}
The mean reflects both how often joint excess occurs and how large it is when it occurs.

In each of 1,999 repetitions, every firm's $N_v$ counts are independently assigned across eligible years using the probabilities in Equation~\eqref{eq:allocation-probability}. Allocations are independent across firms. Assigning more of a firm's fixed total to one year leaves fewer counts for its other years. We calculate excess from the simulated count and the same fitted mean $m_{vt}$ using Equation~\eqref{eq:excess}. Pairs sharing a firm in the same year use the same excess for that firm.

For the upper-tail amount test, the Monte Carlo p-value is one plus the number of simulated $T$ values at least as large as observed, divided by 2,000.

\subsubsection{Results}
\begin{figure}[p]
\centering
\includegraphics[width=0.85\linewidth,height=0.80\textheight,keepaspectratio]{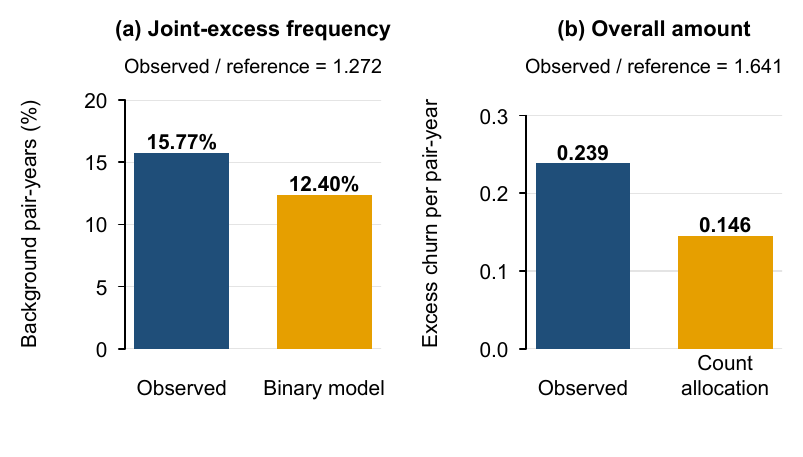}
\caption{Joint excess among firms linked by a transaction in the previous or current year. (a) Share of pair-years in which both firms exceed their fitted mean churn counts, compared with the binary reference matching firm and year exceedance totals in expectation. From 1,999 Bernoulli simulations, upper-tail MC $p=0.0005$ and Holm-adjusted $p=0.001$ across two backgrounds. (b) Overall co-excess amount, including zero pair-years, and the mean from 1,999 independent-firm count allocations. Upper-tail MC $p=0.0005$ and Holm-adjusted $p=0.001$ across the two amount references.}
\label{fig:coexcess}
\end{figure}

Both firms exceed their fitted means in 9.32 million pair-years, compared with 7.33 million expected under the binary reference (Table~\ref{tab:coexcess-occurrence}). The observed count is 27.2\% higher. Figure~\ref{fig:coexcess}(a) shows both counts as shares of all background pair-years.
None of the 1,999 simulated counts reaches the observed count, giving the minimum attainable upper-tail MC $p=0.0005$.

For overall amount, the observed mean is 0.239094 changes per pair-year, compared with 0.145661 under the independent-firm count allocation, a ratio of 1.6414 (Table~\ref{tab:coexcess} and Figure~\ref{fig:coexcess}(b)). None of the 1,999 simulated means reaches the observed value. The upper-tail Monte Carlo $p$ is 0.0005, the minimum attainable value.

As a supplementary check, we compare the overall amount with a reference that randomly reorders each firm's observed excess amounts among its eligible years (Appendix~\ref{app:time-shuffle}). Applying Holm correction across the count-allocation and time-shuffle tests gives an adjusted $p$ of 0.001 for the count-allocation comparison.

Appendix~\ref{app:time-shuffle} decomposes the overall amount into frequency and size under the count-allocation reference.

Year-separated comparisons retain positive alignment on the start-year transaction background (Appendix~\ref{app:joint-multiyear}). From a one-year to a five-year separation, the occurrence ratio decreases from 1.089 to 1.036 and the overall-amount ratio from 1.332 to 1.244. All five comparisons for each measure have Holm-adjusted $p=0.005$ in their respective families. On the identical eligible starting pairs, the matched same-year ratios are larger at every separation: 1.176--1.270 for occurrence and 1.620--1.650 for amount. The stronger contemporaneous alignment therefore coexists with year-separated association.

\subsubsection{Sensitivity to the choice of background}
To assess how the definition of transaction neighbors affects the results, we form the whole-period union $\mathcal U=\bigcup_t\mathcal P_t$ of 9,024,429 distinct pairs. We evaluate each pair in every year with eligible records for both firms, including years without a transaction between them in $t-1$ or $t$. This gives 84,057,287 pair-years. We obtain separate observed and expected counts by applying Equation~\eqref{eq:coexceedance-count} to these pair-years with the same $Z_{vt}$ and $q_{vt}$. We use the same simulated firm-year indicators for both backgrounds in each repetition and apply Holm correction across the two occurrence tests.

With this broader definition, both firms exceed their fitted means in 12.25 million pair-years, 10.1\% above the 11.13 million expected under the binary reference (Table~\ref{tab:coexcess-occurrence}). None of the simulated counts reaches the observed count, giving upper-tail MC $p=0.0005$. Both backgrounds have Holm-adjusted $p=0.001$. Appendix~\ref{app:reproducibility} checks whether the result holds when allowing dependence across years. In the year-separated comparisons, the joint-exceedance count is significantly above its reference only at one- and two-year separations. Overall amount remains above its reference at all five separations (Appendix~\ref{app:joint-multiyear}).

\clearpage

\begin{table}[p]
\centering\small
\caption{Counts of pair-years in which both firms exceed their fitted mean churn counts, compared with the binary reference, 2013--2024. Expected counts sum products of fitted endpoint probabilities. The previous or current year background contains 59,099,713 pair-years. The whole-period union used for the sensitivity check contains 84,057,287 pair-years. Both retain observed zeros, require eligible endpoints in the evaluated year, and use the same 1,999 simulations of firm-year indicators. Holm correction covers the two upper-tail tests.}
\label{tab:coexcess-occurrence}
\begin{tabular}{@{}>{\raggedright\arraybackslash}p{3.4cm}rrrrr@{}}
\toprule
Background & Observed count & Expected count & Ratio & MC $p$ & Holm $p$ \\
\midrule
Previous or current year & 9,321,901 & 7,328,642.768 & 1.271982 & 0.0005 & 0.001 \\
Whole-period union (sensitivity) & 12,251,939 & 11,128,163.230 & 1.100985 & 0.0005 & 0.001 \\
\bottomrule
\end{tabular}
\end{table}

\begin{table}[p]
\centering\small
\caption{Overall co-excess amount against the independent-firm count allocation on 59,099,713 background pair-years. The means include pair-years with zero joint excess. The reference mean uses 1,999 whole-period allocations. Holm correction covers this test and the supplementary shuffle of excess amounts across years within each firm (Appendix~\ref{app:time-shuffle}).}
\label{tab:coexcess}
\begin{tabular}{@{}rrrrr@{}}
\toprule
Observed mean & Count-reference mean & Ratio & MC $p$ & Holm $p$ \\
\midrule
0.239094086 & 0.145661272 & 1.641439 & 0.0005 & 0.001 \\
\bottomrule
\end{tabular}
\end{table}

\FloatBarrier
\subsection{Industry patterns in churn}
\label{sec:industry-patterns}
We describe how firm-level churn varies across industries and years and how joint exceedance varies across industry pairs. Industry is each firm's primary classification in the previous year and can change over time.

\subsubsection{Count concentration across industries}
\begin{figure}[tbp]
\centering
\includegraphics[width=0.85\linewidth,height=0.80\textheight,keepaspectratio]{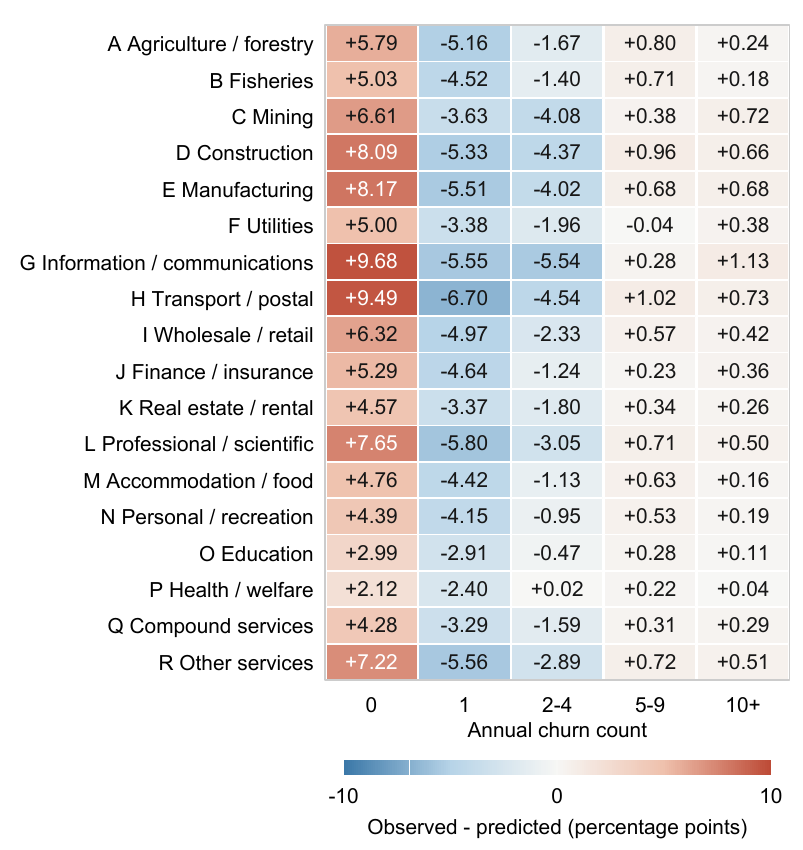}
\caption{Observed minus expected percentages of firm-years in annual gross-churn count ranges by industry, 2013--2024. Red and blue denote positive and negative percentage-point differences. Expected shares use the firm-level count-allocation benchmark in Analysis~1. Every count range in a row uses the same industry-specific firm-year denominator, including zero-churn observations.}
\label{fig:industry-counts}
\end{figure}

Figure~\ref{fig:industry-counts} compares observed and expected shares of firm-years in five annual gross-churn ranges, separately by industry in 2013--2024. Expected shares use the firm-level count-allocation benchmark in Analysis~1 (Section~\ref{sec:temporal-concentration}). Every industry shown has more zero-count years and years with at least ten churn events than expected, and fewer years with exactly one event. Counts of two to four are less frequent than expected in seventeen of the eighteen industries. The exception is health and welfare, with 2.64\% observed versus 2.62\% expected. Information and communications has the largest shortfall in this range, with 13.52\% observed versus 19.06\% expected.

Counts of five to nine are more frequent than expected in seventeen of the eighteen industries shown. Utilities is the exception, with a small deficit of 0.04 percentage points. The positive excess reaches 1.02 points in transport and postal services. The excess of zero-count years ranges from 2.12 points in health and welfare to 9.68 in information and communications. For at least ten events, the observed and expected shares are 1.72\% and 1.06\% in construction, 2.27\% and 1.58\% in manufacturing, and 2.88\% and 1.76\% in information and communications. The magnitude of these differences varies across industries.

\subsubsection{Annual activity and significant upper excess}
\begin{figure}[tbp]
\centering
\includegraphics[width=0.85\linewidth,height=0.80\textheight,keepaspectratio]{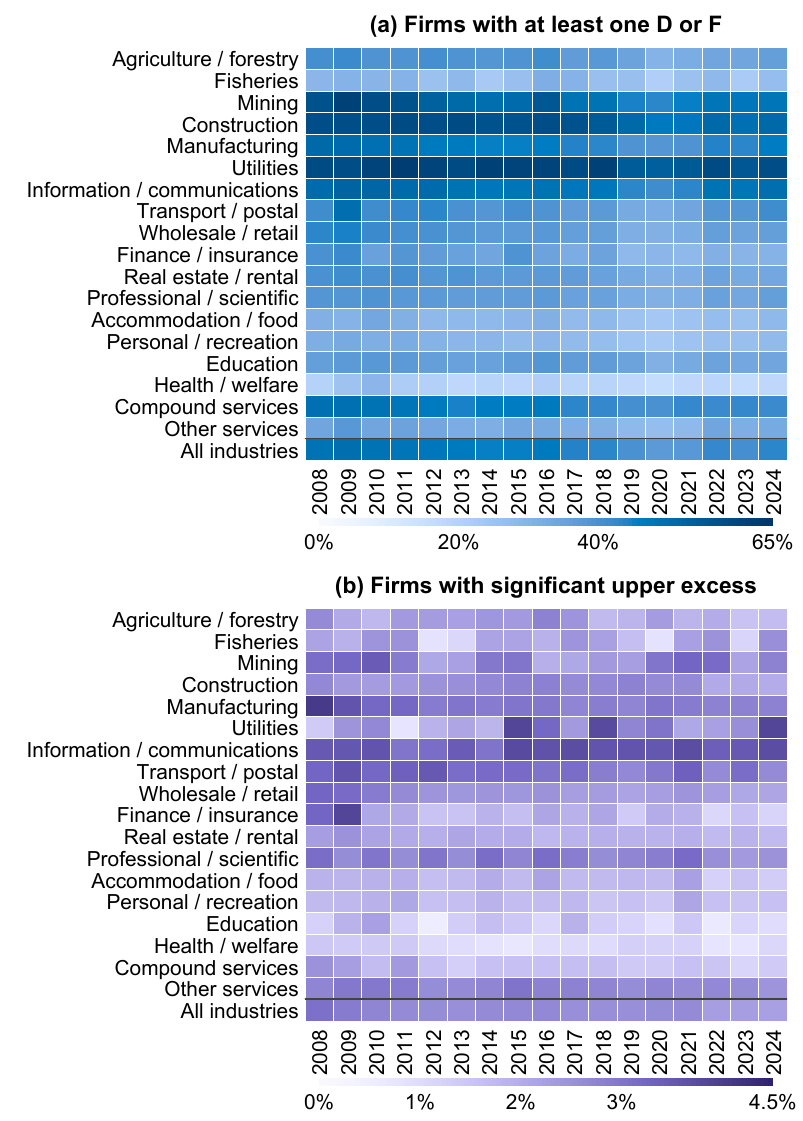}
\caption{Churn by industry and year among 446,328 firms with company records, active status and retained trading relationships throughout 2007--2024. (a) Percentage of firms with any gross churn in each year, 2008--2024. (b) Percentage with significant upper excess under the seventeen-year count-allocation reference and within-firm Holm correction at 5\% across each firm's seventeen years. Industry is the firm's primary classification in the previous year. Each percentage uses the corresponding industry-year firm count; the panels use separate scales.}
\label{fig:industry-year}
\end{figure}

Figure~\ref{fig:industry-year} follows a common cohort of 446,328 firms with company records, active status and at least one retained trading relationship in every snapshot throughout 2007--2024, covering seventeen annual transitions in 2008--2024. Appendix~\ref{app:participation} reports their distribution of years with churn. In Figure~\ref{fig:industry-year}, panel (a) shows the fraction with any incoming or outgoing D or F in each industry-year, and panel (b) shows the fraction whose count significantly exceeds the count-allocation reference for that year. The reference is fitted separately over the full eligible 2008--2024 panel. Upper-tail binomial p-values are Holm-corrected at 5\% across each firm's seventeen years. Each cell uses its industry-year firm count as denominator; the all-industry row uses 446,328.

Activity levels and their timing differ across industries. Across all firms in the cohort, the fraction with churn is highest in 2009 (50.19\%) and lowest in 2020 (37.70\%). Construction ranges from 46.81\% to 59.45\%, compared with 16.14\% to 29.15\% in health and welfare. High activity and above-benchmark activity also have different patterns. Manufacturing, information and communications, and transport and postal services have a higher fraction of significant upper-excess firm-years than the all-industry fraction in every year. In construction, the fraction with any churn peaks in 2011, whereas the significant upper-excess fraction peaks in 2016. Thus raw churn participation alone does not locate the years in which activity most often exceeds firms' own benchmarks.

\FloatBarrier
\subsubsection{Joint excess by industry pair}
\begin{figure}[!b]
\centering
\includegraphics[width=\linewidth]{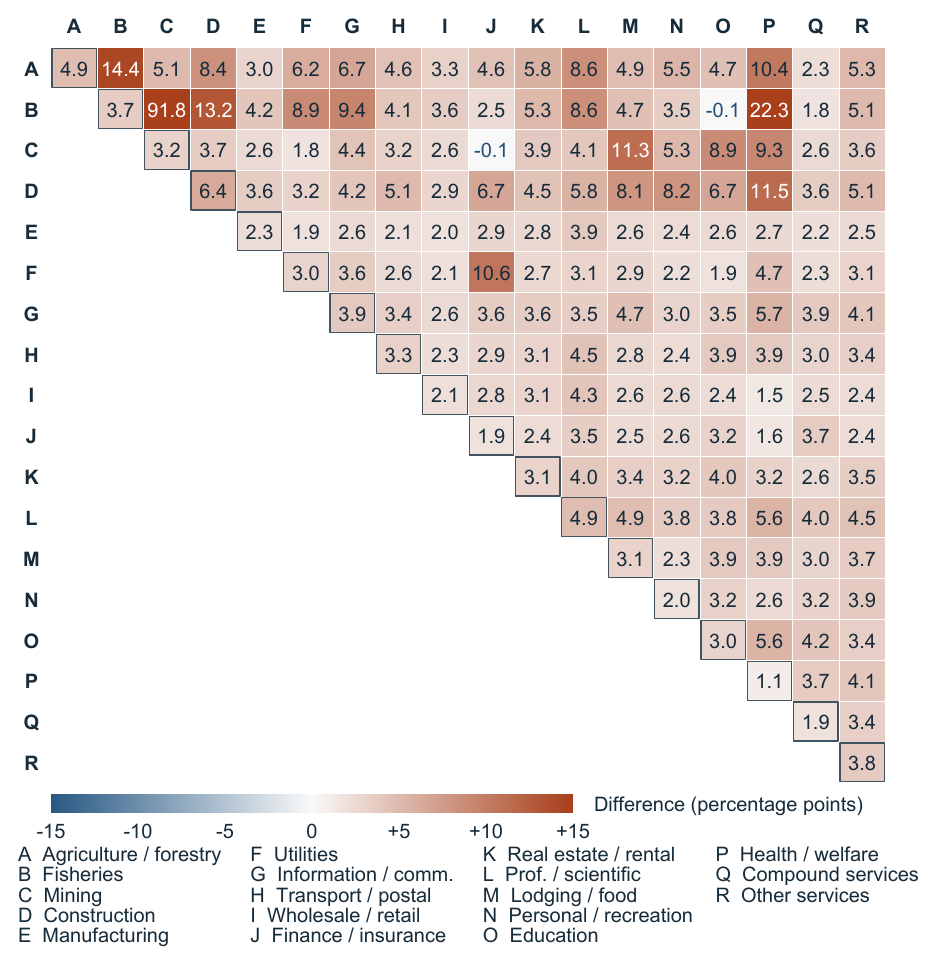}
\caption{Observed minus expected shares of pair-years with joint exceedance by industry pair, 2013--2024, in percentage points. Pairs involving government are excluded. Outlined diagonal cells show same-industry pairs. Colors saturate at $\pm15$ points. The fisheries--mining and fisheries--health and welfare cells contain 2 and 8 pair-years, respectively. The cell differences are descriptive.}
\label{fig:industry-joint-excess}
\end{figure}

We group the pair-years in Analysis~3 (Section~\ref{sec:joint-excess}) by the two firms' industries, regardless of transaction direction. We exclude pair-years with either firm classified as government. Within each industry combination, we divide both the observed and expected counts in Equation~\eqref{eq:coexceedance-count} by the number of pair-years. Figure~\ref{fig:industry-joint-excess} shows the difference between these shares.

The observed share exceeds expectation by 2.93 percentage points across different-industry pair-years and by 3.96 points across same-industry pair-years. It is higher in 151 of the 153 different-industry combinations. For example, construction paired with health and welfare shows a difference of 11.49 points across 58,164 pair-years.

\FloatBarrier
\section{Discussion}
\suppressfloats[t]
Analysis~1 (temporal concentration, Section~\ref{sec:temporal-concentration}) shows that firms have more years with zero or at least five changes, and fewer years with one to four, than the count allocation predicts. Fujii et al. and Kawakubo and Suzuki show that turnover varies with firm characteristics such as age and productivity \cite{fujii2017,kawakubo2022}. Reisch et al. \cite{reisch2026} also document how links form and survive. We hold each firm's full-period churn total fixed and allow fitted activity to vary with year and firm characteristics. The gaps in the annual distribution therefore concern the timing of recorded activity. A model that reproduces overall turnover or link survival can still miss the concentration of changes into bursts.

Concentration and recurrence describe different temporal features. In Appendix~\ref{app:firm-multiyear}, observed next-year gross-churn recurrence is 59.45\%, compared with 60.25\% under the model with firm effects. For D, F, and gross churn, the group with at least ten first-year changes has a higher mean next-year count than its firm-effect prediction. In the common cohort, firms also have fewer pairs of active years than expected at one- to five-year separations for dissolutions and for the restricted set of formations defined in the appendix. These descriptive comparisons distinguish intense episodes from regular repetition. Huneeus \cite{huneeus2018} documents sluggish link adjustment and stronger responses to large firm-specific international trade shocks in Chile. Adjustment costs of the kind he models could produce intermittent changes, while our evidence concerns the observed timing of churn.

Analysis~2 (neighboring churn, Section~\ref{sec:neighboring-churn}) shows that the composition of local adjustment varies by link direction. Within each of the four directions, the relative excess is larger for same-type changes than for mixed types and is largest for F around F. At both endpoints, relative excesses are larger on outgoing links to clients than on incoming links from suppliers. P\'alovics et al. \cite{palovics2021} find more new trading partners among firms exposed to partners' sales shocks than among matched firms. Korovkin et al. \cite{korovkin2026} find different changes in supplier and buyer counts depending on how firms were exposed to conflict. These studies motivate inspecting the supplier and client sides separately. Demir et al. \cite{demir2024} show that demand from richer export destinations is associated with a shift toward skill-intensive domestic partners in Turkey, including among newly formed relationships. The concentration of F around F is consistent with several channels, including expansion and replacement.

Analysis~3 (joint excess churn, Section~\ref{sec:joint-excess}) finds that neighbors exceed their fitted means together 27.2\% more often than predicted after matching firm and year exceedance totals in expectation. The relative excess falls to 10.1\% when we include pairs linked at some point during the period and evaluate them in every eligible year. Matous and Todo study how a firm's existing connections relate to its choice of trading partners, and Asai and Nirei estimate the association between common partners and link formation \cite{matoustodo2016,asainirei2025}. Our outcome concerns the joint timing of churn among trading partners. Under the separate count-allocation reference, the observed mean joint excess is 64.1\% higher than predicted. This mean includes zeros and reflects both how often joint excess occurs and how large it is when it occurs.

Comparisons across years show how far this alignment extends. With the start-year transaction background, the occurrence ratio falls from 1.089 at one year to 1.036 at five years and the amount ratio falls from 1.332 to 1.244 (Appendix~\ref{app:joint-multiyear}). For each separation, the same starting pairs have larger ratios when both outcomes are measured in the starting year. On the whole-period union, occurrence exceeds its reference significantly at separations of one and two years, while amount exceeds its reference at all five. The next-year fixed-total comparisons in Analysis~2 also show a change in composition. Same-type counts remain above expectation and all eight mixed-type means are below expectation in the separately eligible next-year sample (Appendix~\ref{app:multiyear-neighboring}). These results indicate strong same-year alignment alongside some continuing association across years.

Industry patterns show how widely these findings extend. In all eighteen displayed industries, years with zero or at least ten churn events are more common than expected and one-event years are less common (Figure~\ref{fig:industry-counts}). Five-to-nine-event years are also more common except in utilities, and two-to-four-event years are less common except in health and welfare. Churn participation and above-benchmark activity can peak at different times. In construction, the share with any churn peaks in 2011 and the share with significant excess peaks in 2016 (Figure~\ref{fig:industry-year}). Yamano and Sakata find industry differences in partner-set turnover, and Reisch et al. find differences in link entry and exit rates across supplier--customer industry pairs \cite{yamano2023,reisch2026}. Our descriptive pair comparison extends this perspective to above-benchmark activity at both firms. The observed joint-exceedance share is 2.93 percentage points above expectation for different-industry pairs and 3.96 points above expectation for same-industry pairs (Figure~\ref{fig:industry-joint-excess}).

These results complement broader accounts of industrial change. Timmer et al. \cite{timmer2014} trace the international organization of value chains. Gaggl et al. \cite{gaggl2026} distinguish the growth implications of structural change in networks for intermediate inputs and new capital. Our data reveal where and when trading relationships change. Baqaee et al. \cite{baqaee2025} use Belgian data to estimate how changes in supplier access affect downstream marginal costs and use those estimates in growth accounting. Assessing the economic consequences of the rewiring we observe would require information on what firms trade and on their costs or output.

The joint comparisons include shared changes, since one link formation or dissolution contributes to both firms' counts. Shared changes, common conditions and interdependent adjustments can each contribute to the observed alignment. The references generate firm activity on the observed transaction background. The time shuffle preserves firms' observed excess amounts while changing how those amounts line up with annual conditions and the background. The main tests treat the fitted probabilities as fixed, and refitting in the Analysis~1 sensitivity check yields the same reported p-values (Appendix~\ref{app:reference}). The annual snapshots identify which changes occur in the same interval. Their sequence within that interval and the causal mechanisms behind the observed alignment remain open.

\section{Conclusion}
\suppressfloats[t]
Trading relationships among active Japanese firms change in bursts that align in time and around connected firms. For a given full-period churn total, years without changes and large bursts are more common than predicted, while years with one to four changes are less common. Around a changing link, relative excesses are larger for neighboring changes of the same type, with the largest excess for formations on the focal supplier's links to other clients around a formation. Connected firms exceed their fitted means together more often than expected, and their joint excess averaged across all pair-years is greater than the count allocation predicts. The occurrence gap is widespread across industry pairs. Both measures show stronger same-year alignment than comparisons one to five years apart on the same starting pairs.

The comparisons provide empirical targets for models of production-network adjustment. Simulated annual networks can be evaluated on whether they reproduce firm-level bursts, the kinds and directions of nearby changes, and the timing and size of joint adjustments at connected firms. With repeated supplier--client records, the same measurements can help locate years, groups of connected firms, and industry pairs with unusually concentrated rewiring for closer study. Linking those episodes to dated disruptions, transaction values and firm outcomes would provide a basis for investigating their causes and economic consequences.

\clearpage
\appendix
\numberwithin{equation}{section}
\numberwithin{figure}{section}
\numberwithin{table}{section}

\section{Regression specifications and estimates}
\suppressfloats[t]
\label{app:regressions}
Table~\ref{tab:gross-model} reports the gross-count fit used in Analysis~1 (temporal concentration; Section~\ref{sec:temporal-concentration}) and for overall amount in Analysis~3 (joint excess churn; Section~\ref{sec:joint-excess}). The estimation sample differs from the comparison sample because zero-total firms remain at zero in the latter. The adopted model contains five varying controls, firm effects, and year effects, without additional industry effects.

\begin{table}[tbp]
\centering\small
\caption{Gross-churn Poisson regression, 2013--2024. All covariates refer to the previous year. Standard errors are clustered by firm. Fixed-effect coefficients are not displayed.}
\label{tab:gross-model}
\begin{tabular}{@{}>{\raggedright\arraybackslash}p{8.0cm}rrr@{}}
\toprule
Covariate or model detail & Estimate & Standard error & $p$ \\
\midrule
Total degree, $\log(1+x)$ & 0.090 & 0.008 & $<0.001$ \\
Employment, $\log(1+x)$ & 0.129 & 0.015 & $<0.001$ \\
Capital, $\log(1+x)$ & 0.015 & 0.005 & 0.001 \\
Employment missing & 0.560 & 0.055 & $<0.001$ \\
Capital missing & $-0.107$ & 0.038 & 0.005 \\
Firm and year fixed effects & Yes & & \\
Estimation observations & 9,318,333 & & \\
Estimation firms & 953,053 & & \\
Log likelihood & $-12,531,893$ & & \\
\bottomrule
\end{tabular}
\end{table}

Table~\ref{tab:directional-models} gives the four separate count regressions for the same-year comparison in Analysis~2 (neighboring churn; Section~\ref{sec:neighboring-churn}). Outgoing and incoming fits use the corresponding previous degree, with the same other controls as the gross fit. Each fit excludes its own zero-total firms from coefficient estimation and retains them at zero in comparisons. The coefficients describe fitted conditional associations, not causal effects of degree or firm characteristics.

\begin{table}[tbp]
\centering\small
\caption{Poisson regressions for directional D/F counts, 2013--2024. Cells give estimates and firm-clustered standard errors in parentheses. All covariates refer to the previous year.}
\label{tab:directional-models}
\begin{tabular}{@{}>{\raggedright\arraybackslash}p{5.1cm}rrrr@{}}
\toprule
Covariate or model detail & Outgoing D & Incoming D & Outgoing F & Incoming F \\
\midrule
Degree, $\log(1+x)$ & 2.264 (0.006) & 2.573 (0.010) & $-0.830$ (0.024) & $-1.014$ (0.014) \\
Employment, $\log(1+x)$ & $-0.167$ (0.008) & $-0.224$ (0.008) & 0.363 (0.070) & 0.429 (0.028) \\
Capital, $\log(1+x)$ & $-0.056$ (0.006) & $-0.052$ (0.008) & 0.058 (0.014) & 0.097 (0.009) \\
Employment missing & $-0.418$ (0.037) & $-0.775$ (0.051) & 0.946 (0.198) & 1.380 (0.101) \\
Capital missing & $-0.673$ (0.049) & $-0.548$ (0.079) & 0.205 (0.111) & 0.529 (0.074) \\
Degree used & Outgoing & Incoming & Outgoing & Incoming \\
Firm and year effects & Yes & Yes & Yes & Yes \\
Estimation observations & 5,758,570 & 5,796,513 & 6,665,761 & 6,771,022 \\
Estimation firms & 553,596 & 554,820 & 654,706 & 658,654 \\
Log likelihood & $-4,471,278$ & $-3,750,640$ & $-5,714,418$ & $-5,036,586$ \\
\bottomrule
\end{tabular}
\end{table}

\begin{table}[tbp]
\centering\small
\caption{Binary-logit reference for exceeding the gross-count mean, 2013--2024. All covariates refer to the previous year. Firm and year fixed effects are included but their coefficients are not displayed. Constant-zero firms remain in background comparisons with probability zero.}
\label{tab:exceedance-model}
\begin{tabular}{@{}>{\raggedright\arraybackslash}p{10.0cm}r@{}}
\toprule
Covariate or model detail & Estimate or number \\
\midrule
Total degree, $\log(1+x)$ & $-0.011793$ \\
Employment, $\log(1+x)$ & 0.067955 \\
Capital, $\log(1+x)$ & 0.006627 \\
Employment missing & 0.154308 \\
Capital missing & $-0.135347$ \\
Firm and year fixed effects & Yes \\
Finite-estimation observations & 9,302,579 \\
Finite-estimation firms & 937,299 \\
Constant-zero observations retained in comparisons & 2,891,632 \\
Constant-zero firms retained in comparisons & 513,096 \\
Log likelihood & $-5,165,262.015$ \\
\bottomrule
\end{tabular}
\end{table}

Table~\ref{tab:exceedance-model} reports the five slope coefficients of the binary occurrence model in Equation~\eqref{eq:exceedance-logit}. The coefficients describe fitted associations among firms whose exceedance indicators vary over their eligible years. Firms with no exceedances have fitted probabilities of zero and enter the comparison of observed and expected joint exceedances.

\FloatBarrier
\section{Count-frequency tests and reference checks}
\suppressfloats[t]
\label{app:reference}
\subsection{Count-frequency tests and alternative groupings}
The statistic for each listed count or range is its observed firm-year share minus the share predicted by the annual binomial reference. To obtain the expected share for a range, we add the relevant binomial probabilities at each firm-year and average over all 12,194,211 observations. Firms with $N_v=0$ contribute probability one to count zero and zero to the other ranges. We use the same full sample for every comparison, including observed zero counts.

The fine partition separates counts 0 through 9 and pools counts of at least 10. The five ranges shown in Figure~\ref{fig:firm-distribution}(a) are 0, 1, 2--4, 5--9 and 10+. We move the middle boundary in two complete alternative partitions: 0, 1, 2--3, 4--9 and 10+; and 0, 1, 2--5, 6--9 and 10+. The tail-cutoff sensitivity checks instead retain individual counts 0 through 7 with an 8+ tail, or individual counts 0 through 11 with a 12+ tail. The original distribution plots had already been inspected when these additional ranges were fixed. The ranges were fixed before these additional tests, without selecting them from the resulting p-values.

Together these partitions use 21 distinct hypotheses: the twelve exact counts 0 through 11; the three tails 8+, 10+ and 12+; and the six intervals 2--4, 5--9, 2--3, 4--9, 2--5 and 6--9. A range reused in several partitions is tested once. We apply one Holm correction to all 21 raw two-sided p-values, including the sensitivity comparisons; the rows of Tables~\ref{tab:firm-count-frequencies} and~\ref{tab:firm-count-sensitivity} are not separate correction families.

All 1,999 full-history allocations use the same fitted annual probabilities as Analysis~1 (temporal concentration; Section~\ref{sec:temporal-concentration}). For each range, the upper and lower tails count simulated shares at least as large and at least as small as the observed share, respectively. We add one to each tail count and divide by 2,000. The two-sided p-value is twice the smaller tail p-value, capped at one. Its smallest attainable value is 0.001.

All eleven comparisons in Table~\ref{tab:firm-count-frequencies} have the minimum attainable raw two-sided Monte Carlo p-value of 0.001 and a Holm-adjusted p-value of 0.021. Count zero, each individual count from 5 through 9, and the pooled 10+ tail are above expectation; each count from 1 through 4 is below expectation. The additional exact counts 10 and 11 in Table~\ref{tab:firm-count-sensitivity} are also above expectation.

All additional comparisons in Table~\ref{tab:firm-count-sensitivity} likewise have raw two-sided $p=0.001$ and Holm-adjusted $p=0.021$. Changing the middle split to 2--3 and 4--9 gives differences of $-3.153$ and $+0.543$ percentage points; using 2--5 and 6--9 gives $-3.249$ and $+0.639$ percentage points. The 8+ and 12+ tails exceed expectation by 0.838 and 0.342 percentage points. These pooled results retain the lower-range deficit and upper-range excess, while combining counts with different signs: count 4 is below expectation although 4--9 is above, and count 5 is above expectation although 2--5 is below. The integer-specific results therefore remain relevant when interpreting any pooled grouping.

\begin{table}[tbp]
\centering
\caption{Firm-year frequencies in individual churn counts and the pooled upper tail, 2013--2024. Expected shares average the annual binomial probabilities over all 12,194,211 firm-years, retaining observed zeros and firms with zero full-period churn. Difference is observed minus expected, in percentage points. The two-sided Monte Carlo tests use 1,999 fixed-total allocations with the fitted annual probabilities held fixed. Holm correction covers all 21 distinct comparisons across this table and Table~\ref{tab:firm-count-sensitivity}; these eleven rows are not a separate testing family.}
\label{tab:firm-count-frequencies}
\begingroup
\fontsize{10}{12}\selectfont
\setlength{\tabcolsep}{5pt}
\renewcommand{\arraystretch}{1.18}
\begin{tabular}{@{}lrrrrr@{}}
\toprule
Count or range & \shortstack[r]{Observed\\(\%)} & \shortstack[r]{Expected\\(\%)} & \shortstack[r]{Difference\\(pp)} & \shortstack[r]{Two-sided\\MC $p$} & \shortstack[r]{Holm\\$p$} \\
\midrule
0 & 67.404 & 60.224 & $+7.180$ & 0.001 & 0.021 \\
1 & 13.956 & 19.070 & $-5.114$ & 0.001 & 0.021 \\
2 & 6.450 & 8.743 & $-2.294$ & 0.001 & 0.021 \\
3 & 3.687 & 4.546 & $-0.859$ & 0.001 & 0.021 \\
4 & 2.384 & 2.552 & $-0.168$ & 0.001 & 0.021 \\
5 & 1.583 & 1.510 & $+0.072$ & 0.001 & 0.021 \\
6 & 1.099 & 0.934 & $+0.166$ & 0.001 & 0.021 \\
7 & 0.779 & 0.601 & $+0.178$ & 0.001 & 0.021 \\
8 & 0.562 & 0.402 & $+0.161$ & 0.001 & 0.021 \\
9 & 0.413 & 0.278 & $+0.134$ & 0.001 & 0.021 \\
10+ & 1.684 & 1.141 & $+0.543$ & 0.001 & 0.021 \\
\bottomrule
\end{tabular}
\endgroup
\end{table}

\begin{table}[tbp]
\centering
\caption{Additional count ranges used for Figure~\ref{fig:firm-distribution}(a) and alternative groupings. The groups distinguish the extra integer counts and upper tails used for the tail-cutoff sensitivity checks, the pooled ranges shown in the main figure, and the alternative middle boundaries. The remaining ranges needed for each complete partition appear in Table~\ref{tab:firm-count-frequencies}. All rows use the same sample, statistic, fixed-probability Monte Carlo procedure and single Holm family of 21 comparisons. Each distinct range appears once across the two tables.}
\label{tab:firm-count-sensitivity}
\begingroup
\fontsize{10}{12}\selectfont
\setlength{\tabcolsep}{5pt}
\renewcommand{\arraystretch}{1.18}
\begin{tabular}{@{}lrrrrr@{}}
\toprule
Count or range & \shortstack[r]{Observed\\(\%)} & \shortstack[r]{Expected\\(\%)} & \shortstack[r]{Difference\\(pp)} & \shortstack[r]{Two-sided\\MC $p$} & \shortstack[r]{Holm\\$p$} \\
\midrule
\multicolumn{6}{@{}l}{\emph{Tail cutoff sensitivity}} \\[3pt]
10 & 0.310 & 0.199 & $+0.111$ & 0.001 & 0.021 \\
11 & 0.237 & 0.147 & $+0.090$ & 0.001 & 0.021 \\
8+ & 2.659 & 1.820 & $+0.838$ & 0.001 & 0.021 \\
12+ & 1.136 & 0.794 & $+0.342$ & 0.001 & 0.021 \\
\midrule
\multicolumn{6}{@{}l}{\emph{Pooled ranges in Figure 2(a)}} \\[3pt]
2--4 & 12.520 & 15.841 & $-3.321$ & 0.001 & 0.021 \\
5--9 & 4.435 & 3.724 & $+0.711$ & 0.001 & 0.021 \\
\midrule
\multicolumn{6}{@{}l}{\emph{Middle boundary sensitivity}} \\[3pt]
2--3 & 10.136 & 13.289 & $-3.153$ & 0.001 & 0.021 \\
4--9 & 6.819 & 6.276 & $+0.543$ & 0.001 & 0.021 \\
2--5 & 14.103 & 17.351 & $-3.249$ & 0.001 & 0.021 \\
6--9 & 2.853 & 2.214 & $+0.639$ & 0.001 & 0.021 \\
\bottomrule
\end{tabular}
\endgroup
\end{table}

\FloatBarrier
\subsection{Whole-distribution diagnostic and simulated reference}
As a supplementary whole-distribution diagnostic, we measure the maximum absolute difference between the observed cumulative count distribution and its analytic reference over all integer counts. We calculate this distance for each of the 1,999 allocations against the same analytic reference. The largest observed cumulative-frequency difference is 7.18 percentage points, at zero churn. None of the simulated distances reaches this value, giving the minimum attainable upper-tail Monte Carlo p-value, $p=0.0005$. This global diagnostic is separate from the 21 count-frequency hypotheses and does not assign significance to individual counts or ranges.

\begin{figure}[tbp]
\centering
\includegraphics[width=\linewidth,height=0.80\textheight,keepaspectratio]{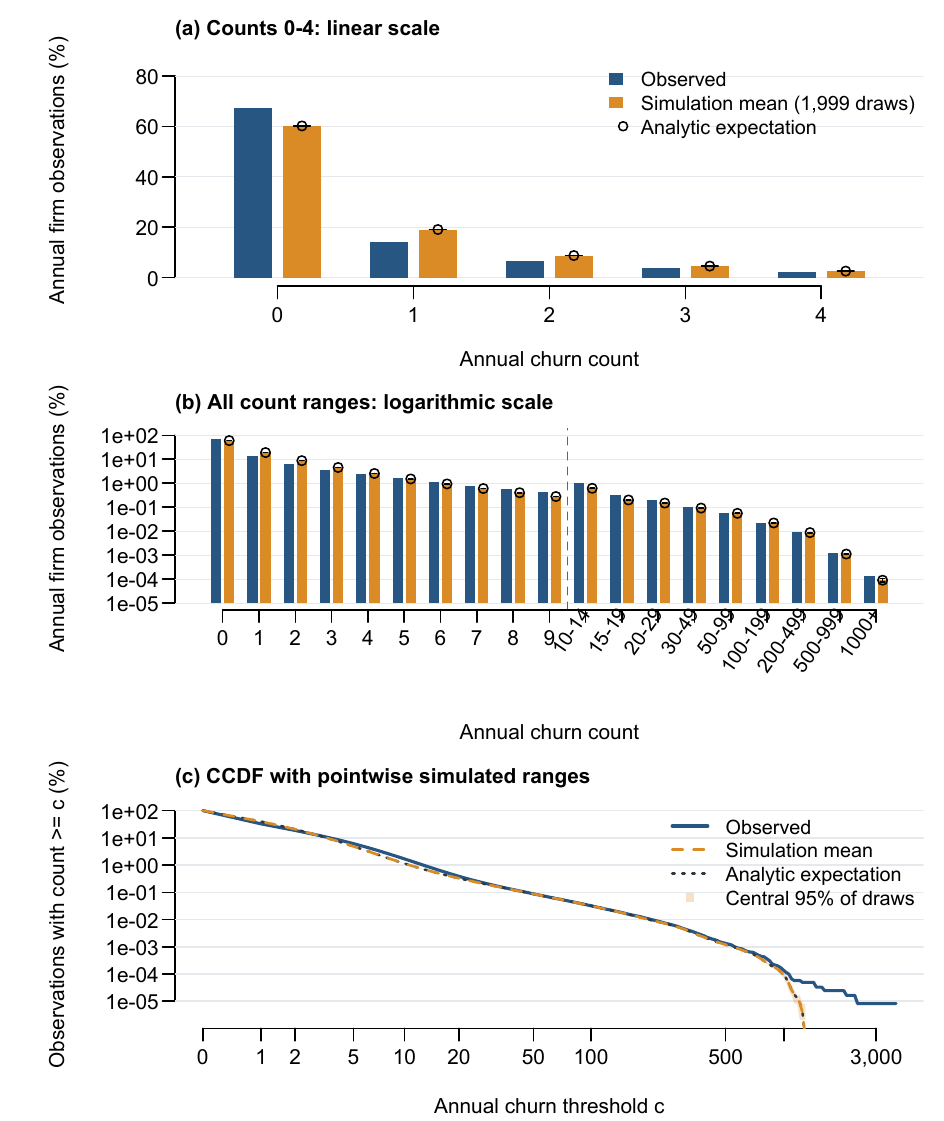}
\caption{Fixed-parameter simulation reference for annual gross counts in Figure~\ref{fig:firm-distribution}. (a) Counts 0--4. (b) All count ranges, with a logarithmic vertical axis; the dashed line separates single-count and wider bins. Bars are shares in each range, not densities. (c) Complementary cumulative distributions using $\log(1+c)$ on the horizontal axis to include zero. Open circles give analytic expectations; simulated ranges are pointwise central 95\% ranges from 1,999 draws, without refitting. Every denominator includes all 12,194,211 firm-years.}
\label{fig:firm-reference}
\end{figure}

Figure~\ref{fig:firm-reference} compares the analytic expectation from the firm-level count allocation with the mean of 1,999 fixed-parameter allocations. Every draw keeps each firm's total and the number of observations fixed. The pooled mean is therefore 1.109046 in both observations and every draw. The simulated mean closely matches the analytic expectation, reproducing the excess observed zeros and heavy upper tail relative to either reference construction.

The central ranges in Figure~\ref{fig:firm-reference} show pointwise variation in shares from the same allocations used for the tests.

As a sensitivity check, we refit the count model to each of the same 1,999 datasets and compare each simulated distribution against its own fitted expectation. All 21 comparisons yield the same two-sided and Holm-adjusted Monte Carlo p-values, 0.001 and 0.021. The supplementary whole-distribution test also yields the same upper-tail p-value, 0.0005.

\FloatBarrier
\section{Neighboring count distributions}
\begin{figure}[tbp]
\centering
\includegraphics[width=\linewidth,height=0.80\textheight,keepaspectratio]{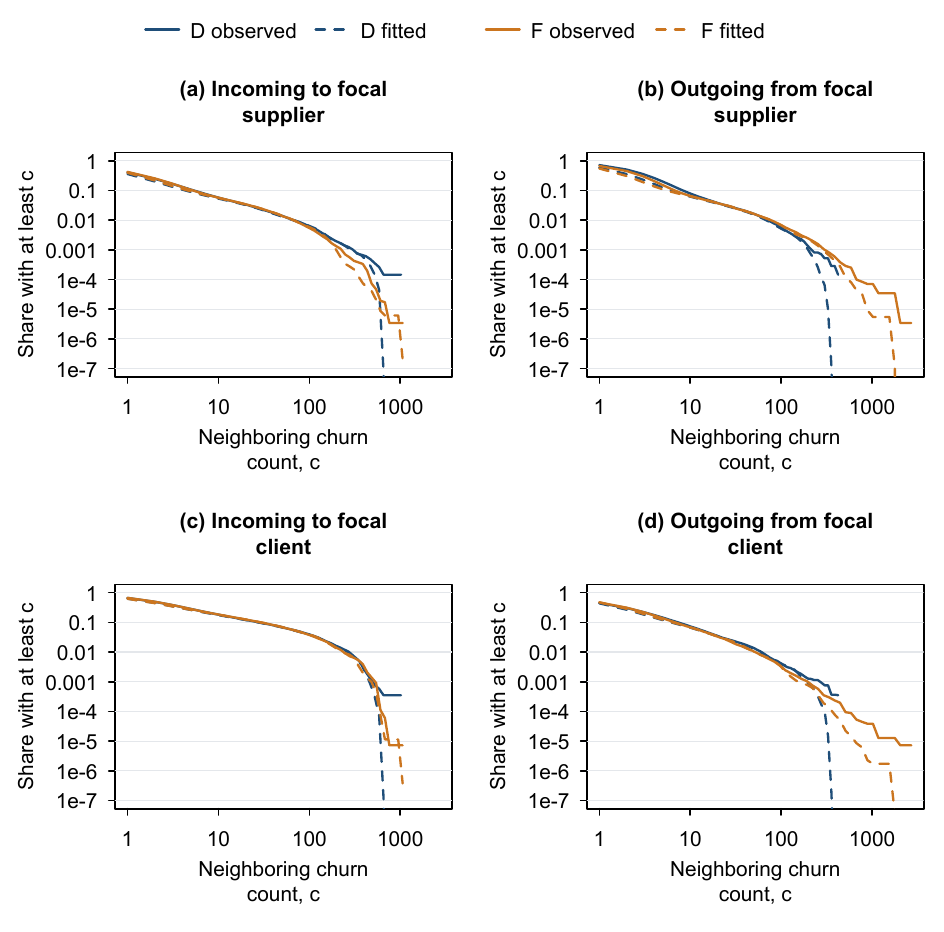}
\caption{Neighboring D/F count distributions around the 2,917,266 observed focal D, 2013--2024. Solid lines show observed fractions and dashed lines fitted probabilities. Panels separate the directions in Figure~\ref{fig:directions}. Both axes are logarithmic, and zero counts remain in the denominator. The focal D is excluded from matching outgoing-supplier and incoming-client D counts.}
\label{fig:neighbor-d}
\end{figure}

\begin{figure}[tbp]
\centering
\includegraphics[width=\linewidth,height=0.80\textheight,keepaspectratio]{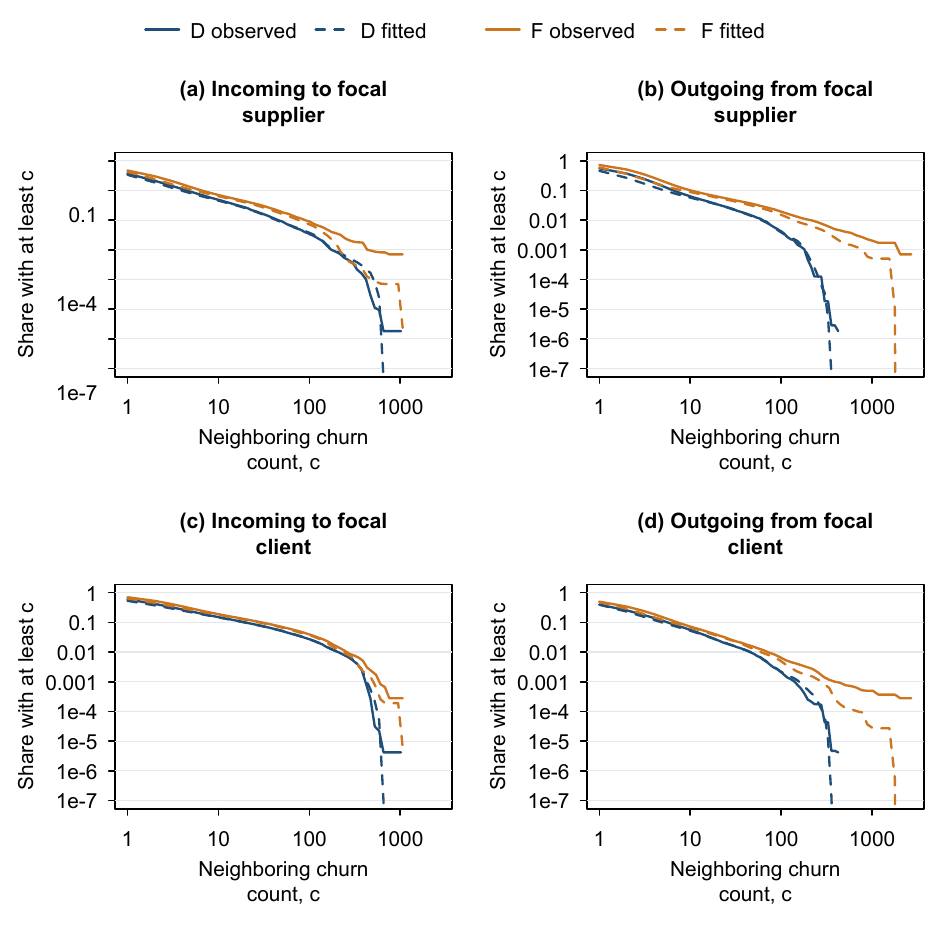}
\caption{Neighboring D/F count distributions around the 3,844,702 observed focal F. Definitions and scales follow Figure~\ref{fig:neighbor-d}. The focal F is excluded from matching outgoing-supplier and incoming-client F counts.}
\label{fig:neighbor-f}
\end{figure}

\suppressfloats[t]
\label{app:neighbor-distributions}
Table~\ref{tab:neighbors} reports the observed and expected mean neighboring counts for Analysis~2 (neighboring churn; Section~\ref{sec:neighboring-churn}). Figures~\ref{fig:neighbor-d} and \ref{fig:neighbor-f} give the complementary cumulative distributions underlying the mean comparisons in Table~\ref{tab:neighbors}, separately for focal D and F. Each curve includes all focal churn events of that type in the denominator, even if there are zero neighboring changes. The benchmark averages analytic binomial probabilities over the same observed focal events as the corresponding observed curve.

The curves show why a positive mean difference does not imply an excess at every threshold. In fifteen of sixteen comparisons, observations have fewer zero counts and more counts of at least ten than expected, but some curves cross at larger counts. For focal F and neighboring F on the focal supplier's other outgoing links, 1.93\% have at least 100 neighboring formations, compared with 1.52\% expected. The adopted tests concern overall mean differences, not the individual curve points.

\begin{table}[tbp]
\centering\small
\caption{Mean same-year neighboring counts around focal D/F, 2013--2024. Entries give observed / expected counts. Column labels give focal / neighboring type. Each comparison has two-sided $p=0.001$ and Holm-adjusted $p=0.016$ across sixteen comparisons, from 1,999 fixed-fit allocations.}
\label{tab:neighbors}
\begin{tabular}{@{}lrrrr@{}}
\toprule
Direction & D/D & D/F & F/D & F/F \\
\midrule
\texttt{in\_supplier} & 3.65 / 3.28 & 3.37 / 3.17 & 2.55 / 2.45 & 5.24 / 3.73 \\
\texttt{out\_supplier} & 4.74 / 3.77 & 4.70 / 4.03 & 3.57 / 3.10 & 13.01 / 7.91 \\
\texttt{in\_client} & 14.46 / 13.74 & 14.05 / 13.63 & 10.66 / 10.50 & 15.16 / 13.37 \\
\texttt{out\_client} & 3.65 / 3.24 & 3.36 / 3.17 & 2.56 / 2.47 & 5.24 / 3.63 \\
\bottomrule
\end{tabular}
\end{table}

\FloatBarrier
\section{Supplementary co-excess comparisons}
\suppressfloats[t]
\label{app:time-shuffle}
\subsection{Frequency and size under the count reference}
\begin{figure}[tbp]
\centering
\includegraphics[width=\linewidth,height=0.80\textheight,keepaspectratio]{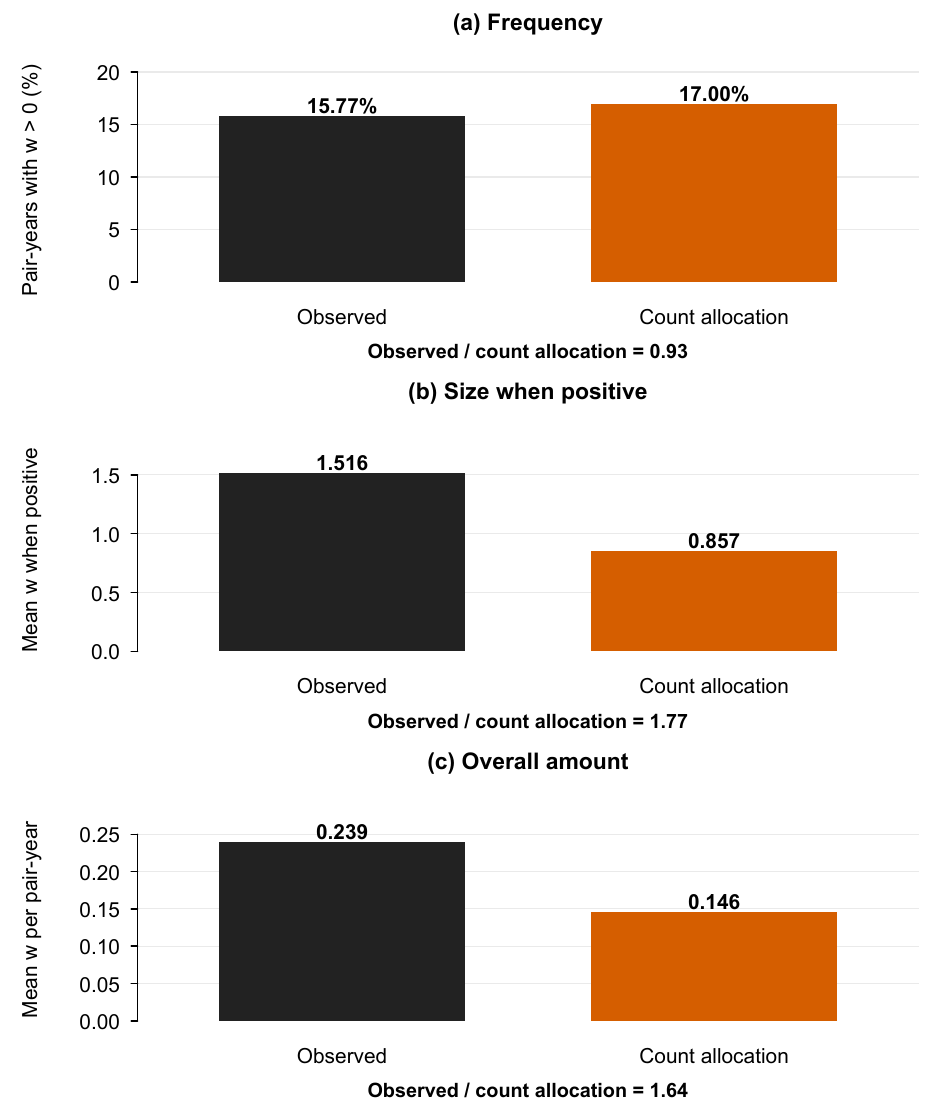}
\caption{Frequency, conditional mean size, and overall amount of positive co-excess relative to the firm-level count allocation for pairs linked in the previous or current year. The reference uses 1,999 independent-firm, whole-period allocations. Only overall amount has the reported test. These components use the count reference throughout, rather than the binary occurrence reference used in panel (a) of Figure~\ref{fig:coexcess}.}
\label{fig:coexcess-count-components}
\end{figure}

The count-allocation reference used for overall amount also implies an occurrence frequency, but it does not match firms' observed exceedance counts. Both firms exceed their fitted mean churn counts in 15.773175\% of observed pair-years and 16.995745\% under that reference. Among pair-years with positive joint excess, the mean amount is 1.515827 observed and 0.857046 expected. The frequency ratio of 0.928066 multiplied by the conditional-size ratio of 1.768666 gives the overall-amount ratio of 1.641439 (Figure~\ref{fig:coexcess-count-components}). The reference mean when joint excess occurs is the expected total amount divided by the expected number of pair-years with positive joint excess.

Under the count reference, joint excess occurs less often in the observations, but the amount when it occurs is larger. Table~\ref{tab:coexcess-occurrence} compares occurrence against a binary reference fitted directly to the exceedance indicator.

\subsection{Annual count-reference comparisons}
\begin{figure}[tbp]
\centering
\includegraphics[width=\linewidth,height=0.80\textheight,keepaspectratio]{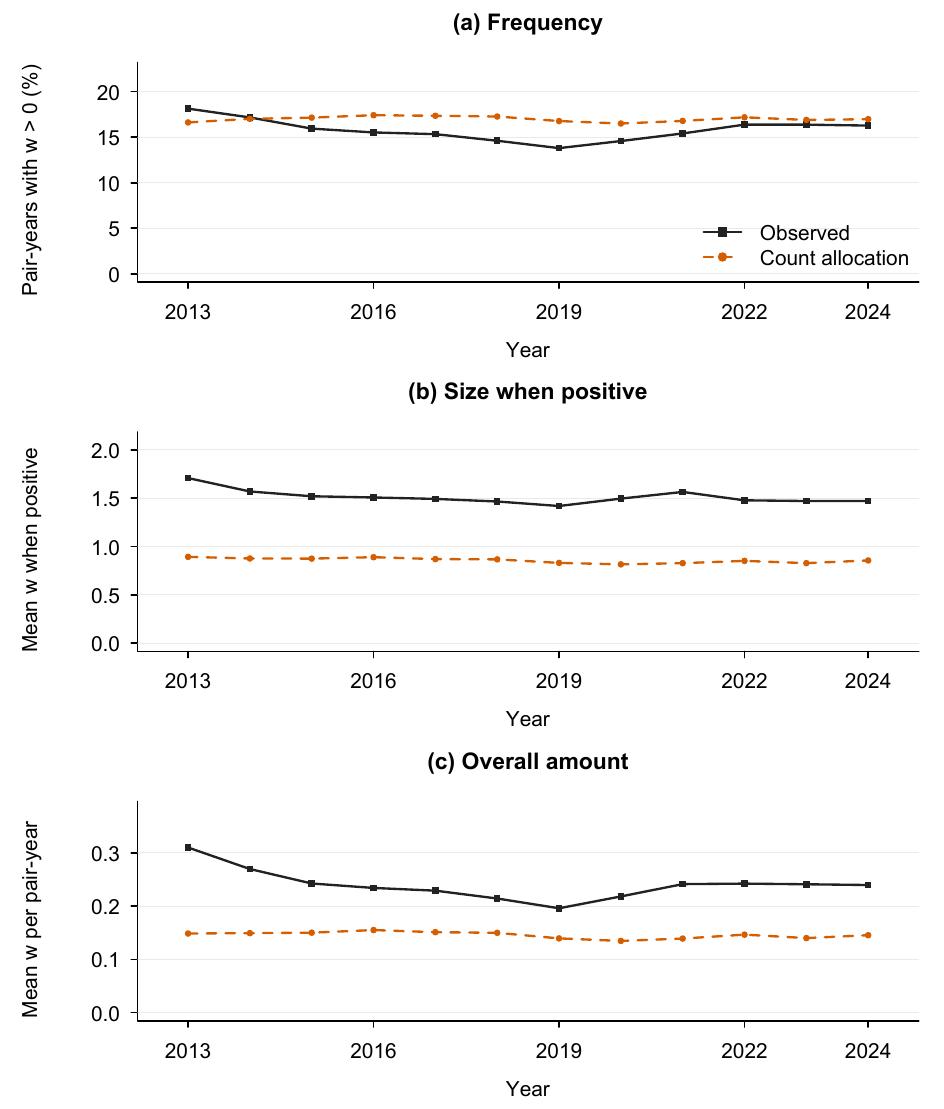}
\caption{Annual positive co-excess at transaction neighbors relative to the firm-level count allocation, 2013--2024. The panels use the same definition of joint excess and count reference as Figure~\ref{fig:coexcess-count-components}. Within each year, the mean under the reference when joint excess occurs is the expected total amount divided by the expected number of pairs with positive joint excess. The annual comparisons are descriptive.}
\label{fig:annual-coexcess}
\end{figure}

Figure~\ref{fig:annual-coexcess} displays frequency, conditional size, and overall amount separately by year under the count reference on the same background as the pooled result. The annual comparisons are descriptive.

\subsection{Observed-excess time shuffle}
The supplementary time shuffle asks how the overall joint amount changes when every firm's observed positive-excess values are retained but their years are independently permuted. We move the excess values themselves, including zeros, across each firm's eligible years. We do not move counts and rescore them against the destination year's mean. The same permuted firm-year value is used on every incident background pair. The observed annual backgrounds remain fixed separately in each year. This preserves the firm's entire excess multiset while changing its annual matching with other firms and with its background position.

The reference assumes uniformly exchangeable eligible-year orderings conditional on those multisets and the fixed background. It does not preserve population-wide annual activity or within-firm serial order. Because the background varies by year, it also changes the correspondence between an excess value and the firm's degree, position, and number of incident background pairs. The resulting comparison does not isolate interfirm alignment alone.

\begin{figure}[tbp]
\centering
\includegraphics[width=\linewidth,height=0.80\textheight,keepaspectratio]{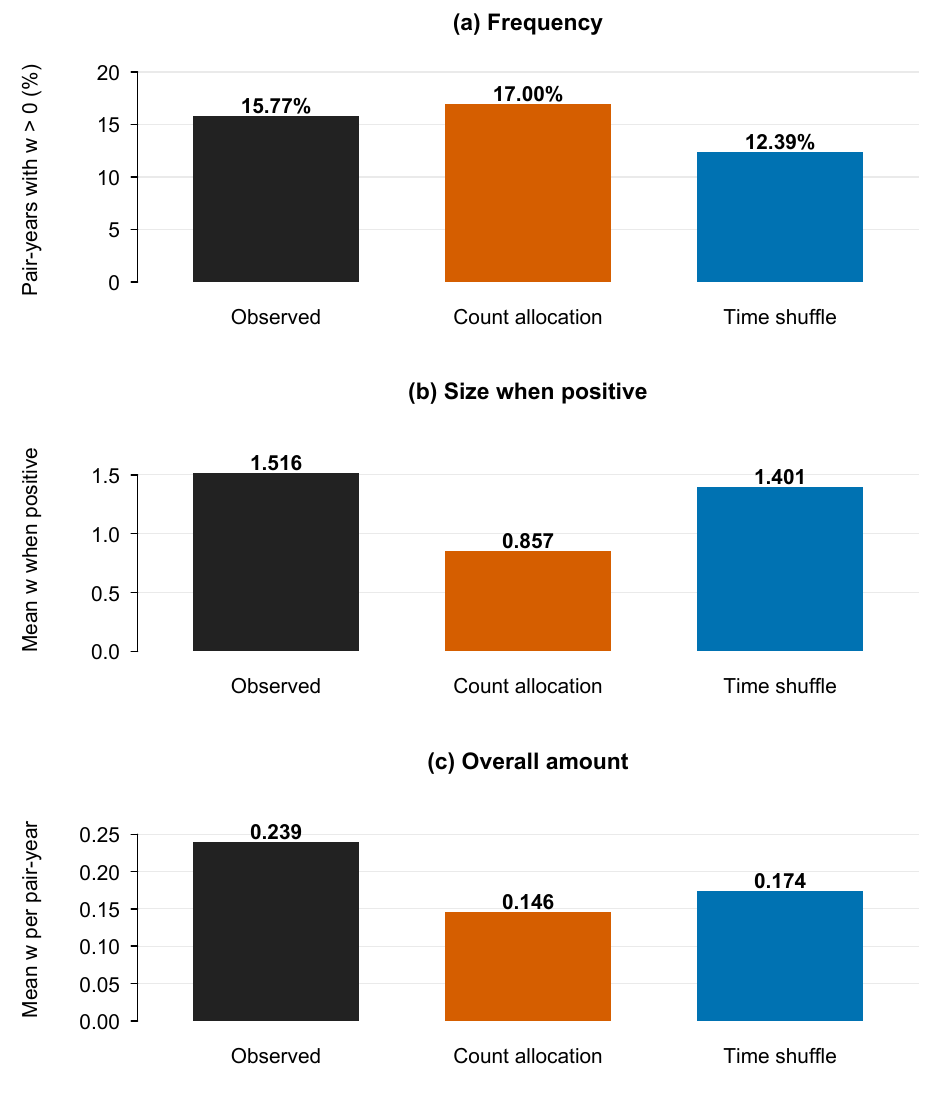}
\caption{Co-excess frequency, conditional size, and overall amount under the count allocation and observed-excess time shuffle. Both references use 1,999 draws and the same background pair-years. The conditional mean under each reference is the expected total amount divided by the expected number of pair-years with positive joint excess.}
\label{fig:time-shuffle}
\end{figure}

The observed overall mean of 0.239094 is 1.3778 times the shuffle mean of 0.173537. None of the 1,999 draws reaches the observed value, giving upper-tail $p=0.0005$ and Holm-adjusted $p=0.001$ across the two overall-amount tests. Joint excess occurs in 15.773\% of observed pair-years versus 12.387\% under the shuffle. When it occurs, the mean amounts are 1.51583 and 1.40097, respectively (Figure~\ref{fig:time-shuffle}). In an exploratory comparison added after examining the amount results, the frequency difference under the shuffle has upper-tail MC $p=0.0005$ and Holm-adjusted $p=0.002$. That Holm correction covers four comparisons, one for frequency and one for overall amount under each of the count and shuffle references. Conditional-size summaries are descriptive.

\FloatBarrier
\section{Churn participation in a common cohort}
\begin{figure}[tbp]
\centering
\includegraphics[width=\linewidth,height=0.80\textheight,keepaspectratio]{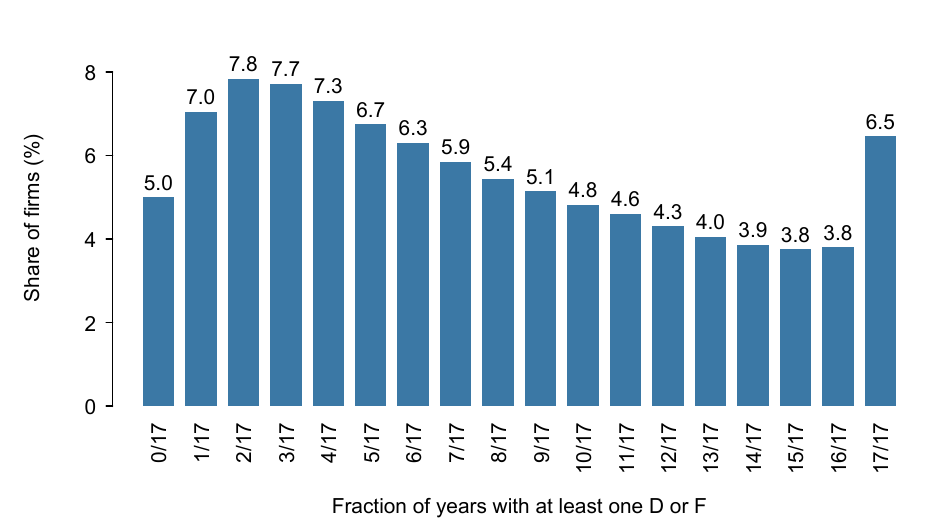}
\caption{Fraction of years with churn among the common cohort of 446,328 firms. Each bar gives the percentage with gross churn in exactly $k$ of the seventeen transitions, at position $k/17$. Each firm is counted once. A year with churn has at least one D or F, not necessarily a statistically significant upper excess.}
\label{fig:active-years}
\end{figure}

\suppressfloats[t]
\label{app:participation}
The main sample assesses eligibility year by year. To describe a fixed population over a longer period, Figure~\ref{fig:active-years} instead follows the 446,328 firms with company records, active status and at least one retained trading relationship in every snapshot throughout 2007--2024. It counts the number of the seventeen transitions, 2008--2024, with any incoming or outgoing D or F. This is a count of years with churn, not a significance test or a direct measure of serial dependence.

\FloatBarrier
\section{Sensitivity to dependence across years}
\suppressfloats[t]
\label{app:reproducibility}
The joint-exceedance test in Analysis~3 (Section~\ref{sec:joint-excess}) assumes independence across years. Here we allow arbitrary dependence across years while keeping firm indicators within each year independent at their fitted probabilities $q_{vt}$.

For each background, let $A_t$ be the joint-exceedance count in year $t$ and $A=\sum_t A_t$ the pooled count. The variance of $A_t$ accounts for pairs sharing a firm. Without specifying how the annual counts are related, Cauchy--Schwarz bounds the pooled variance by $\operatorname{Var}(A)\leq V_{\max}=(\sum_t\sqrt{\operatorname{Var}(A_t)})^2$. For the observed excess $\Delta=A_{\mathrm{obs}}-\mathbb E A>0$, Cantelli's inequality bounds the probability of reaching the observed count by $\Pr(A-\mathbb E A\geq\Delta)\leq V_{\max}/(V_{\max}+\Delta^2)$.

After Holm correction across the two backgrounds, these upper bounds are approximately 0.00434 for the previous/current background and 0.01311 for the whole-period union. Both are below 5\%, so the joint-exceedance result holds under this relaxation of independence across years.

\FloatBarrier
\section{Network size and firm-level churn statistics}
\begin{figure}[!bp]
\centering
\includegraphics[width=\linewidth,height=0.80\textheight,keepaspectratio]{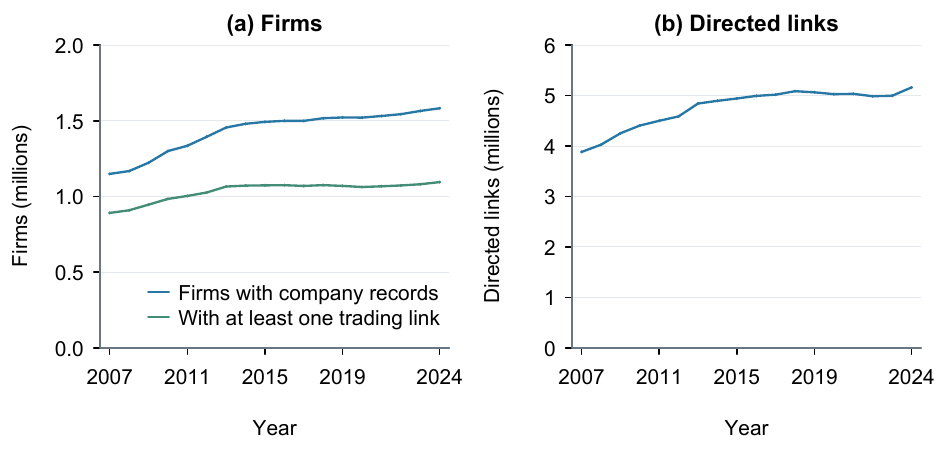}
\caption{Annual numbers of firms and directed links, 2007--2024. Panel (a) includes firms with company records and active status in each year, and the subset with at least one trading link. Panel (b) counts unique directed links between these firms.}
\label{fig:annual-network}
\end{figure}

\suppressfloats[t]
\label{app:annual-network}
Figure~\ref{fig:annual-network} and Table~\ref{tab:annual-network} report annual snapshot counts for 2007--2024. Firms have company records and active status in the reported year. A linked firm has at least one incoming or outgoing link to another firm in this set; links count unique directed supplier--client pairs. These yearly totals describe the available network before applying the consecutive-year linked-firm eligibility rule for churn.

\begin{table}[!bp]
\centering\small
\caption{Annual snapshot counts of firms and directed links, 2007--2024. Definitions follow Figure~\ref{fig:annual-network}.}
\label{tab:annual-network}
\begin{tabular}{@{}rrrr@{}}
\toprule
Year & \shortstack{Firms with\\company records} & \shortstack{With\\trading links} & Directed links \\
\midrule
2007 & 1,148,904 & 891,308 & 3,881,778 \\
2008 & 1,167,628 & 908,692 & 4,024,654 \\
2009 & 1,222,461 & 945,929 & 4,250,698 \\
2010 & 1,299,700 & 983,737 & 4,404,322 \\
2011 & 1,334,571 & 1,003,307 & 4,498,751 \\
2012 & 1,394,064 & 1,026,148 & 4,585,910 \\
2013 & 1,455,168 & 1,065,379 & 4,840,624 \\
2014 & 1,479,770 & 1,071,619 & 4,893,959 \\
2015 & 1,492,449 & 1,073,214 & 4,940,272 \\
2016 & 1,499,283 & 1,074,362 & 4,991,850 \\
2017 & 1,498,964 & 1,069,551 & 5,016,405 \\
2018 & 1,516,663 & 1,074,800 & 5,084,929 \\
2019 & 1,521,542 & 1,069,620 & 5,062,472 \\
2020 & 1,520,605 & 1,062,331 & 5,026,275 \\
2021 & 1,531,127 & 1,067,123 & 5,032,593 \\
2022 & 1,543,222 & 1,072,600 & 4,985,316 \\
2023 & 1,564,586 & 1,080,607 & 4,994,776 \\
2024 & 1,582,404 & 1,094,852 & 5,159,103 \\
\bottomrule
\end{tabular}
\end{table}

\FloatBarrier
\begin{table}[p]
\centering\small
\caption{Descriptive statistics of gross churn per firm-year in the main 2013--2024 panel. All observations, including zeros, are retained.}
\label{tab:firm-churn-statistics}
\begin{tabular}{@{}>{\raggedright\arraybackslash}p{8.8cm}r@{}}
\toprule
Statistic & Value \\
\midrule
Annual firm observations & 12,194,211 \\
Distinct firms & 1,450,395 \\
Observations with zero churn & 8,219,441 (67.40\%) \\
Mean churn count per firm per year & 1.109 \\
Standard deviation & 5.532 \\
Minimum & 0 \\
25th percentile & 0 \\
Median & 0 \\
75th percentile & 1 \\
90th percentile & 3 \\
95th percentile & 5 \\
99th percentile & 12 \\
99.9th percentile & 45 \\
Maximum & 3,794 \\
\bottomrule
\end{tabular}

\par\vspace{10pt}
\noindent\begin{minipage}{\linewidth}
\normalsize\raggedright
Table~\ref{tab:firm-churn-statistics} summarizes annual gross churn, defined in Equation~\eqref{eq:gross}, in the main 2013--2024 panel. Each observation is a firm-year, including zeros; the same firm can contribute observations in multiple eligible years. Each D/F churn event enters both endpoint firms' counts.

Percentiles are the smallest integer counts whose cumulative proportions reach the stated percentages.
\end{minipage}
\end{table}

\FloatBarrier
\section{Multiyear patterns of firm-level churn}
\label{app:firm-multiyear}
Extending Analysis~1 (temporal concentration; Section~\ref{sec:temporal-concentration}), we examine whether churn continues at the same firm beyond a single annual interval. Dissolution (D) combines incoming and outgoing D, formation (F) combines incoming and outgoing F, and gross churn is their sum. These are firm-level histories, not repeated changes of the same individual relationship. The recurrence and next-year count comparisons use the eligible 2008--2024 panel: 16,666,885 firm-years from 1,618,430 firms. Their 16 adjacent-year comparisons contain 14,889,394 observations, with first years 2008--2023 and following years 2009--2024. Each comparison requires eligibility in both annual intervals, hence active company records and retained trading relationships in all three snapshots. A firm need not appear throughout the full period.

\subsection{Next-year recurrence and usual firm activity}
For each churn measure, activity means that its annual count is positive. We fit separate logistic models using the logarithm of one plus previous total incoming-plus-outgoing degree and year effects, first without and then with firm effects. The latter allow persistent differences in firms' activity. Neither specification includes lagged churn as a regressor. Constant binary histories receive boundary probabilities of zero or one in the firm-effect specification and remain in the comparison.

Let $A_{vt}$ indicate activity and $r_{vt}$ be the fitted activity probability for the measure being examined. Let $\mathcal C$ contain the eligible firm and first-year combinations. Conditional independence of activity across years at these probabilities gives the fitted recurrence rate
\begin{equation}
R^{\mathrm{obs}}=
\frac{\sum_{(v,t)\in\mathcal C}A_{vt}A_{v,t+1}}
     {\sum_{(v,t)\in\mathcal C}A_{vt}},\qquad
R^{\mathrm{fit}}=
\frac{\sum_{(v,t)\in\mathcal C}r_{vt}r_{v,t+1}}
     {\sum_{(v,t)\in\mathcal C}r_{vt}}.
\label{eq:firm-recurrence-rate}
\end{equation}
The fitted rate is the ratio of expected overlap to expected first-year activity. It is not the mean next-year probability among only the firms observed to be active in the first year. Year effects reproduce full-panel annual activity totals in expectation; firm effects additionally reproduce each firm's number of active years in expectation, rather than fixing it in every realization.

Table~\ref{tab:firm-recurrence-benchmarks} shows that observed recurrence exceeds the degree-and-year prediction for all three measures. With firm effects, the predictions instead exceed the observed rates: gross recurrence is 59.45\%, compared with 47.26\% without and 60.25\% with firm effects. The same probabilities can be combined over all seventeen years for the 446,328 firms eligible throughout 2007--2024. In that common cohort, 159,013 firms have gross churn in at least ten years, compared with approximately 130,631 and 162,816 under the two specifications. These are in-sample fitted comparisons; no validated tests or uncertainty intervals for their differences are available. Their ordering does not establish an absence of temporal dependence.

\subsection{Continuation after different first-year counts}
We retain each observed first-year count group, 0, 1, 2--4, 5--9, or at least 10. Groups are defined separately from D, F, and gross counts, so they need not contain the same observations across measures. The zero group concerns subsequent onset, not recurrence. Within a group, the next-year activity prediction is the mean of the fitted next-year logistic probabilities, with no first-year probability weighting.

For next-year amounts, we use separate Poisson models with the same degree-and-year specifications, again with and without firm effects. Firms with zero total counts retain zero means in the firm-effect comparison. Table~\ref{tab:firm-next-year-counts} averages the observed and fitted next-year counts over the same observations within each first-year group, including zeros. The last column subtracts the firm-effect prediction from the observed mean.

Adding firm effects reduces the gaps, but the at-least-ten groups retain positive differences: 0.215 D, 0.767 F, and 0.214 gross churn events per firm-year. Their next-year activity proportions also exceed the firm-effect predictions by 3.26, 2.32, and 1.16 percentage points. The one-count groups have negative activity gaps of 4.21, 3.56, and 2.81 points in magnitude, and both the one-count and 2--4-count groups have negative mean-count gaps for all three measures. High-count continuation therefore coexists with different patterns after modest churn. The fits and groups use the same histories, so these gaps remain descriptive. The Poisson D benchmark is not bounded by feasible previous links: its means exceed eligible degree in 15,709 observations without and 8,907 with firm effects.

\subsection{Activity one to five years apart under fixed totals}
A separate comparison follows the common cohort of 446,328 firms eligible throughout 2007--2024. It examines the seventeen annual changes in 2008--2024 and uses a different allocation benchmark, rather than the logistic or Poisson regressions above. Incoming and outgoing events are combined within D or F. Formation events in years with zero eligible prior degree are omitted from both the fixed firm totals and the observed activity measure; this removes 4,851 events. Eligible degree counts previous links whose two endpoints qualify for the annual transition.

For each type, we retain firms with positive full-period event totals and positive eligible degree in at least one year: 387,712 for D and 413,640 for F. Years with zero events remain. Initial annual weights multiply each firm's exact eligible degree by the pooled event rate among these retained firms in its year and degree group. Positive-degree groups are 1, 2--3, 4--7, and successive powers-of-two ranges through 16,384--32,767 and 32,768 or more. We multiply these weights by common year multipliers, normalize within each firm, and iteratively update the multipliers to match annual event totals in expectation. Each firm's observed total remains fixed exactly in the allocation model.

Write $u_{vt}$ for the resulting annual allocation probabilities and $N_v^{*}$ for the retained event total. Under a multinomial allocation of that total, the probability of activity in both years separated by $\ell$ is
\begin{equation}
\pi_{v,t,\ell}
=1-(1-u_{vt})^{N_v^{*}}-(1-u_{v,t+\ell})^{N_v^{*}}
 +(1-u_{vt}-u_{v,t+\ell})^{N_v^{*}}.
\label{eq:firm-lag-overlap}
\end{equation}
We sum observed binary overlaps and Equation~\eqref{eq:firm-lag-overlap} over firms and all eligible year combinations at each separation. Ten events in a year count no more than one in this statistic. Expectations are calculated analytically, without random draws.

At separations of one to five years, observed overlaps are 73.1--74.7\% of the reference for D and 74.4--75.6\% for the restricted F measure (Table~\ref{tab:firm-lag-overlap}). These are descriptive fixed-total comparisons without a network-dependence-calibrated test. The lower binary overlap under this allocation and the positive high-count gaps above concern different samples, activity measures, and references; neither establishes that churn causes subsequent churn or that temporal dependence is absent.

\begin{table}[p]
\centering\small
\caption{Next-year recurrence in the eligible linked-firm panel, pooling 16 adjacent-year comparisons in 2008--2024. The observed rate conditions on first-year activity. Fitted rates are ratios of expected counts in Equation~\eqref{eq:firm-recurrence-rate}. Rates are percentages. These are descriptive in-sample comparisons.}
\label{tab:firm-recurrence-benchmarks}
\begin{tabular}{@{}>{\raggedright\arraybackslash}p{1.5cm}rrrr@{}}
\toprule
Measure & \shortstack{Observed first-year\\active observations} & \shortstack{Observed\\recurrence (\%)} & \shortstack{Degree + year\\prediction (\%)} & \shortstack{+ firm\\prediction (\%)} \\
\midrule
D & 3,214,805 & 45.44 & 36.82 & 49.53 \\
F & 4,061,641 & 48.95 & 38.10 & 52.12 \\
Gross & 5,262,030 & 59.45 & 47.26 & 60.25 \\
\bottomrule
\end{tabular}
\end{table}

\begin{table}[p]
\centering\small
\caption{Next-year mean counts by observed first-year count in the eligible linked-firm panel. Each measure's five groups total 14,889,394 adjacent-year observations. D and F combine incoming and outgoing changes. Each measure defines its own first-year groups. Zero first-year counts measure onset. All group comparisons are descriptive.}
\label{tab:firm-next-year-counts}
\begin{tabular}{@{}lrrrrrr@{}}
\toprule
Measure & \shortstack{First-year\\count} & \shortstack{Adjacent-year\\observations} & \shortstack{Observed\\mean} & \shortstack{Degree + year\\mean} & \shortstack{+ firm\\mean} & \shortstack{Observed\\minus + firm} \\
\midrule
D & 0 & 11,674,589 & 0.258 & 0.319 & 0.251 & +0.007 \\
D & 1 & 1,757,881 & 0.694 & 0.604 & 0.726 & -0.032 \\
D & 2--4 & 1,150,705 & 1.313 & 0.987 & 1.350 & -0.037 \\
D & 5--9 & 238,966 & 2.955 & 2.295 & 2.933 & +0.021 \\
D & 10+ & 67,253 & 17.726 & 16.717 & 17.511 & +0.215 \\
F & 0 & 10,827,753 & 0.335 & 0.424 & 0.333 & +0.002 \\
F & 1 & 2,068,115 & 0.788 & 0.753 & 0.809 & -0.022 \\
F & 2--4 & 1,525,662 & 1.409 & 1.204 & 1.433 & -0.024 \\
F & 5--9 & 360,673 & 3.022 & 2.549 & 3.043 & -0.020 \\
F & 10+ & 107,191 & 17.811 & 15.110 & 17.044 & +0.767 \\
Gross & 0 & 9,627,364 & 0.462 & 0.665 & 0.461 & +0.001 \\
Gross & 1 & 2,195,895 & 1.094 & 1.096 & 1.108 & -0.014 \\
Gross & 2--4 & 2,044,390 & 1.902 & 1.598 & 1.933 & -0.030 \\
Gross & 5--9 & 741,214 & 3.638 & 2.756 & 3.655 & -0.016 \\
Gross & 10+ & 280,531 & 16.410 & 14.265 & 16.196 & +0.214 \\
\bottomrule
\end{tabular}
\end{table}

\begin{table}[p]
\centering\small
\caption{Activity in both years at one- to five-year separations under the common linked-cohort fixed-total allocation. Incoming and outgoing events are combined within type. Formation omits events in zero-eligible-degree years. These comparisons are descriptive.}
\label{tab:firm-lag-overlap}
\begin{tabular}{@{}lrrrr@{}}
\toprule
Measure & \shortstack{Separation\\(years)} & Observed & Expected & Ratio \\
\midrule
D & 1 & 1,059,021 & 1,417,750.9 & 0.747 \\
D & 2 & 984,274 & 1,320,782.2 & 0.745 \\
D & 3 & 903,682 & 1,222,647.1 & 0.739 \\
D & 4 & 832,480 & 1,133,691.7 & 0.734 \\
D & 5 & 764,231 & 1,046,092.0 & 0.731 \\
F & 1 & 1,387,887 & 1,836,029.8 & 0.756 \\
F & 2 & 1,292,552 & 1,713,193.0 & 0.754 \\
F & 3 & 1,199,841 & 1,597,679.1 & 0.751 \\
F & 4 & 1,106,717 & 1,479,747.3 & 0.748 \\
F & 5 & 1,018,894 & 1,370,010.9 & 0.744 \\
\bottomrule
\end{tabular}
\end{table}
\FloatBarrier

\FloatBarrier
\section{Neighboring churn across years}
\suppressfloats[t]
\label{app:multiyear-neighboring}
Extending Analysis~2 (neighboring churn; Section~\ref{sec:neighboring-churn}), we examine whether churn around a focal churn event continues into the next year, and whether earlier neighboring churn is associated with the probability of a focal D/F churn event. The first two comparisons hold observed focal D/F fixed and evaluate next-year activity at its endpoints. The last uses directed firm pairs with and without a focal D/F churn event to model focal D/F itself. These are different observation units and benchmarks; none identifies a causal sequence of responses.

\subsection{Next-year counts under a fixed-total allocation}
\label{app:neighbor-next-allocation}
The next-year counterpart of Analysis~2 (neighboring churn; Section~\ref{sec:neighboring-churn}) retains the four directions around the focal supplier and client and the same controlled count fits. Both focal firms must have active company records and at least one retained trading link in each of the previous, focal, and following snapshots. There are 2,625,807 focal D and 3,382,797 focal F in 2013--2023, followed by neighboring counts in 2014--2024. We count each focal churn event once, including those followed by zero neighboring churn.

The four directional D/F fits are those used in the same-year comparison, with the count-model and annual-allocation construction in Equations~\eqref{eq:poisson}--\eqref{eq:allocation-probability}. Each firm's total for a directional D/F channel is fixed over 2013--2024. For focal churn event $r$ in year $t$, let $N_v^{(b,d)}$ be its endpoint's full-period total for next-year outcome type $b$ and direction $d$, and let $p_{v,t+1}^{(b,d)}$ be its fitted next-year allocation probability. The next-year reference is
\begin{equation}
Y_{r,t+1}^{(b,d)}\sim\operatorname{Binomial}\!\left(N_v^{(b,d)}-\delta_r^{(b,d)},\ p_{v,t+1}^{(b,d)}\right).
\label{eq:neighbor-next-allocation}
\end{equation}
Here $\delta_r^{(b,d)}=1$ only when the focal churn event has type $b$ and occupies the focal supplier's outgoing or the focal client's incoming channel; otherwise it is zero. This removes the focal churn event already assigned to year $t$ from the total available for year $t+1$. Nothing is subtracted from the observed next-year count. The focal relationship can reappear as F after D or disappear as D after F, and those opposite-type changes are included in both observations and predictions. Consecutive D/D or F/F cannot occur on that same relationship under the annual definitions and therefore concern other links.

We average the analytic next-year distributions over the same observed focal churn events. The mean comparisons use 1,999 fixed-fit joint allocations of the four count channels. Each allocation retains the original partner-eligibility histories attached to directional incidences. All incidence groups contribute to neighboring counts, and the focal-event average uses the simulated counts from slots that satisfy the next-year endpoint-eligibility rule. Each allocation retains firms' directional full-period totals; it does not generate a complete transaction network. Two-sided Monte Carlo probabilities are twice the smaller finite-simulation tail probability, capped at one, with Holm correction over sixteen comparisons. Parameters are held fixed, and the mean tests do not cover every distributional threshold.

Table~\ref{tab:neighbor-next-allocation} shows selective continuation. D after D exceeds the reference in all four directions by 2.0--3.8\%, and F after F by 6.2--24.5\%. The largest excess is outgoing F at the focal supplier after focal F: 8.142 observed versus 6.539 expected. All eight mixed-type comparisons are below the reference. 16 of the sixteen mean differences remain significant at 5\% after Holm correction. The global maximum-absolute-relative-difference test gives the minimum attainable upper-tail Monte Carlo $p=0.0005$. The broad positive pattern in the same-year comparison therefore becomes selective by type in the following year. The next-year comparison additionally requires endpoint eligibility in the following snapshot, so the contrast is not an isolated effect of changing the time interval.

\subsection{Next-year directional occurrence and counts}
\label{app:neighbor-next-regression}
A separate comparison uses the full linked-firm panel in 2008--2024: 16,666,885 firm-years from 1,618,430 firms. Sixteen regressions predict incoming and outgoing D/F separately, using logistic models for occurrence and Poisson models for counts, each with previous total incoming-plus-outgoing degree and year effects, with and without firm effects. Neither previous churn nor focal churn enters as a predictor. The degree variable is $\log(1+k)$; these models do not contain the employment and capital controls of the controlled allocation models. Structural D zeros at zero corresponding previous degree are retained with probability and mean zero. Constant histories have the appropriate boundary predictions in the firm-effect models.

Observed focal D/F in 2008--2023 receive the next-year endpoint-eligibility rule: both focal firms must have active company records and retained trading links in all three snapshots. The resulting 3,779,596 focal D and 5,282,100 focal F give 9,061,696 focal churn events, with next-year outcomes in 2009--2024. We average the fitted next-year probabilities and means over these observed events. Opposite-type changes on the focal relationship are retained. This comparison uses the same linked-firm eligibility condition, a longer period and different reference models from the fixed-total comparison above.

Table~\ref{tab:neighbor-next-regression} reports observations and the references including firm effects. The findings depend on both the direction and the measured outcome. For outgoing F at the focal supplier after F, observed occurrence is 51.25\% versus 50.82\% predicted, while the mean count is 8.957 versus 8.088, an excess of 0.870. Occurrence and count differences therefore describe distinct aspects of next-year churn. These are descriptive in-sample comparisons: the models were estimated from the same histories, and no validated significance tests or confidence intervals for these selected-sample gaps are reported.

\subsection{Contemporaneous and distributed-lag link regressions}
\label{app:neighbor-distributed-lag}
The link-level regressions reverse the roles of outcome and neighboring counts. Separate sampling-weighted linear probability models predict focal D and F in 2013--2024. Both endpoints must have active company records and retained trading links in the previous and current snapshots. The dissolution risk set contains previously existing links, including survivors. The formation risk set contains previously unlinked ordered firm pairs, including pairs that remain unlinked. All observed events and sampled non-events enter with annual weights. The fitted samples contain 17,503,644 rows for D and 23,068,212 for F, representing 56,953,491 existing-link intervals and 12,393,778,458,891 unlinked-pair intervals, respectively.

For focal outcome $Y^{(o)}_{ijt}$, neighboring counts $X^{(m,d)}_{ijt,k}$ record type $m\in\{D,F\}$ in one of the four directions $d\in\mathcal D$ in Figure~\ref{fig:directions} at lag $k$, excluding the focal relationship. Lag 0 is the interval $t-1$ to $t$; lag $k\geq1$ is $t-k-1$ to $t-k$. The outcome always remains a one-year change. Each outcome model jointly includes all 48 neighboring counts, rather than fitting each lag separately:
\begin{equation}
\begin{split}
Y^{(o)}_{ijt}={}&\sum_{k=0}^{5}\sum_{m\in\{D,F\}}\sum_{d\in\mathcal D}
\beta^{(o)}_{mdk}X^{(m,d)}_{ijt,k}+\mathbf W_{ijt}^{\mathsf T}\boldsymbol\gamma^{(o)}\\
&+\alpha^{(S,o)}_i+\alpha^{(C,o)}_j+\tau^{(o)}_t+\varepsilon^{(o)}_{ijt}.
\end{split}
\label{eq:neighbor-lag-lpm}
\end{equation}
The controls $\mathbf W_{ijt}$ include supplier-outgoing and client-incoming degrees measured at $t-6$, endpoint eligibility indicators for the five prior intervals, and the number of neighboring links for each direction, type and lag. These ten indicators flag whether each endpoint has active company records and retained trading links in both snapshots of each earlier interval; they distinguish unavailable history from eligible zero churn. Degrees and link numbers enter as $\log(1+x)$. The last numbers use the start of the interval for D and the end for F. Supplier-role, client-role and year fixed effects allow persistent endpoint heterogeneity and annual conditions, while standard errors are clustered by supplier and client. These effects do not hold all firm-specific annual conditions fixed. The formation population includes many pairs that are not realistic trading opportunities, so its coefficients describe associations in that statistical risk set.

Tables~\ref{tab:neighbor-lag-supplier}--\ref{tab:neighbor-lag-client} report all 48 neighboring-count coefficients for each outcome, in percentage points per additional event. An additional same-year D on another outgoing link of the focal supplier is associated with a 0.2495-point increase in focal-D probability (standard error 0.0406). For neighboring D entering the focal supplier and leaving the focal client, models including lags 1--5 but excluding lag 0 give five-lag sums of $0.04410$ ($p=0.00225$) and $0.01987$ points ($p=0.00128$). Adding lag 0 gives sums of $-0.00725$ ($p=0.719$) and $-0.02903$ ($p=0.142$), respectively. Each sum represents one additional event at every tested lag, not one isolated event.

The two lag specifications estimate different associations: lagged terms conditional on contemporaneous churn are residual associations, whereas omitting contemporaneous churn also retains associations related to same-year activity. Lag sums and joint tests also answer different questions: for outgoing-client D, the sum test gives $p=0.142$ and the joint test of all five coefficients gives $p=7.75\times10^{-9}$. For neighboring F entering the supplier, the conditional five-lag sum in the focal-D model is $-0.01468$ points ($p=4.83\times10^{-5}$). For focal F, none of the eight five-lag sums or joint tests is significant at the nominal 5\% level. For focal D, 7 of eight joint tests reject zero, but all 4 significant five-lag sums are negative. Individual coefficient and lag-sum probabilities are unadjusted for the many outcomes, directions and lags examined. These comparisons do not distinguish causal responses from continuing reorganization or other common conditions.

\begin{table}[tbp]
\centering\fontsize{10}{12}\selectfont
\caption{Next-year neighboring counts under the retained fixed-total reference. Focal years are 2013--2023 and outcome years 2014--2024. Each row uses all 2,625,807 focal D or 3,382,797 focal F. Means include zeros; excess is $(\mathrm{observed}/\mathrm{expected}-1)\times100$. The last column gives the two-sided Holm-adjusted probability across sixteen mean comparisons, from 1,999 fixed-fit allocations.}
\label{tab:neighbor-next-allocation}
\begin{tabular}{@{}cc>{\raggedright\arraybackslash}p{3.5cm}rrrr@{}}
\toprule
Focal & Next & Position & Observed & Expected & Excess (\%) & Holm $p$ \\
\midrule
D & D & Supplier: incoming & 3.219 & 3.154 & +2.07 & 0.016 \\
D & F & Supplier: incoming & 3.113 & 3.128 & -0.50 & 0.016 \\
D & D & Supplier: outgoing & 3.411 & 3.285 & +3.85 & 0.016 \\
D & F & Supplier: outgoing & 4.037 & 4.113 & -1.84 & 0.016 \\
D & D & Client: incoming & 13.626 & 13.361 & +1.98 & 0.016 \\
D & F & Client: incoming & 13.750 & 13.885 & -0.97 & 0.016 \\
D & D & Client: outgoing & 3.179 & 3.093 & +2.78 & 0.016 \\
D & F & Client: outgoing & 3.069 & 3.147 & -2.48 & 0.016 \\
F & D & Supplier: incoming & 2.572 & 2.665 & -3.49 & 0.016 \\
F & F & Supplier: incoming & 3.954 & 3.337 & +18.48 & 0.016 \\
F & D & Supplier: outgoing & 3.487 & 3.593 & -2.95 & 0.016 \\
F & F & Supplier: outgoing & 8.142 & 6.539 & +24.51 & 0.016 \\
F & D & Client: incoming & 10.901 & 11.124 & -2.00 & 0.016 \\
F & F & Client: incoming & 13.489 & 12.703 & +6.19 & 0.016 \\
F & D & Client: outgoing & 2.496 & 2.567 & -2.79 & 0.016 \\
F & F & Client: outgoing & 3.869 & 3.309 & +16.94 & 0.016 \\
\bottomrule
\end{tabular}
\end{table}

\begin{table}[tbp]
\centering\fontsize{10}{12}\selectfont
\caption{Next-year directional occurrence and mean counts in the linked-firm panel. Observed focal years are 2008--2023 and outcomes 2009--2024. References include degree, year and firm effects. Occurrence columns give percentages from the logistic reference; count columns give means from the separate Poisson reference. Rows use 3,779,596 focal D or 5,282,100 focal F, including opposite-type churn on the focal relationship. No tests of these gaps are reported.}
\label{tab:neighbor-next-regression}
\begin{tabular}{@{}cc>{\raggedright\arraybackslash}p{3.4cm}rrrr@{}}
\toprule
& & & \multicolumn{2}{c}{Occurrence (\%)} & \multicolumn{2}{c}{Mean count} \\
Focal & Next & Position & Observed & Predicted & Observed & Predicted \\
\midrule
D & D & Supplier: incoming & 35.47 & 34.13 & 3.706 & 3.552 \\
D & F & Supplier: incoming & 39.02 & 38.88 & 3.843 & 3.804 \\
D & D & Supplier: outgoing & 45.52 & 44.84 & 4.168 & 3.924 \\
D & F & Supplier: outgoing & 52.05 & 50.16 & 4.976 & 4.895 \\
D & D & Client: incoming & 57.96 & 59.12 & 14.682 & 14.358 \\
D & F & Client: incoming & 62.43 & 61.68 & 15.530 & 15.454 \\
D & D & Client: outgoing & 39.69 & 39.16 & 3.692 & 3.523 \\
D & F & Client: outgoing & 43.85 & 43.62 & 3.867 & 3.899 \\
F & D & Supplier: incoming & 35.09 & 36.60 & 3.177 & 3.310 \\
F & F & Supplier: incoming & 41.13 & 39.48 & 4.679 & 4.014 \\
F & D & Supplier: outgoing & 47.15 & 45.95 & 4.558 & 4.379 \\
F & F & Supplier: outgoing & 51.25 & 50.82 & 8.957 & 8.088 \\
F & D & Client: incoming & 58.33 & 56.14 & 11.700 & 11.787 \\
F & F & Client: incoming & 58.70 & 59.39 & 14.764 & 13.726 \\
F & D & Client: outgoing & 36.58 & 37.75 & 2.986 & 3.081 \\
F & F & Client: outgoing & 42.95 & 41.62 & 4.458 & 3.889 \\
\bottomrule
\end{tabular}
\end{table}

\begin{table}[p]
\centering\fontsize{10}{12}\selectfont
\caption{Distributed-lag coefficients for neighboring churn at the focal supplier. Entries are coefficient (standard error), both in percentage points per additional neighboring event. Supplier and client tables partition the same jointly estimated model: each focal outcome includes all 48 neighboring counts, controls and supplier-role, client-role and year effects. Standard errors use supplier/client two-way clustering. Coefficient probabilities are not adjusted for multiple comparisons.}
\label{tab:neighbor-lag-supplier}
\begin{tabular}{@{}>{\raggedright\arraybackslash}p{3.2cm}ccrr@{}}
\toprule
Position & Neighbor & Lag & Focal D & Focal F \\
\midrule
Supplier: incoming & D & 0 & $-0.0583\;(0.0232)$ & $-1.19\times10^{-6}\;(2.36\times10^{-6})$ \\
Supplier: incoming & D & 1 & $0.0068\;(0.014)$ & $1.52\times10^{-6}\;(1.48\times10^{-6})$ \\
Supplier: incoming & D & 2 & $-9.60\times10^{-4}\;(0.0067)$ & $-1.98\times10^{-6}\;(2.32\times10^{-6})$ \\
Supplier: incoming & D & 3 & $1.22\times10^{-4}\;(0.0066)$ & $1.10\times10^{-6}\;(1.04\times10^{-6})$ \\
Supplier: incoming & D & 4 & $-0.0032\;(0.0067)$ & $6.04\times10^{-7}\;(8.00\times10^{-7})$ \\
Supplier: incoming & D & 5 & $-0.0099\;(0.0058)$ & $-2.13\times10^{-6}\;(1.73\times10^{-6})$ \\
\noalign{\vskip 3pt}
Supplier: incoming & F & 0 & $0.0272\;(0.008)$ & $1.80\times10^{-5}\;(6.85\times10^{-6})$ \\
Supplier: incoming & F & 1 & $-0.0108\;(0.0019)$ & $-8.14\times10^{-6}\;(4.25\times10^{-6})$ \\
Supplier: incoming & F & 2 & $-0.0024\;(0.0021)$ & $5.03\times10^{-6}\;(5.94\times10^{-6})$ \\
Supplier: incoming & F & 3 & $-6.12\times10^{-4}\;(0.0013)$ & $-3.00\times10^{-6}\;(3.94\times10^{-6})$ \\
Supplier: incoming & F & 4 & $-3.66\times10^{-4}\;(0.002)$ & $6.28\times10^{-8}\;(1.98\times10^{-6})$ \\
Supplier: incoming & F & 5 & $-5.10\times10^{-4}\;(0.0018)$ & $4.36\times10^{-6}\;(2.97\times10^{-6})$ \\
\midrule
Supplier: outgoing & D & 0 & $0.2495\;(0.0406)$ & $-9.09\times10^{-6}\;(8.11\times10^{-6})$ \\
Supplier: outgoing & D & 1 & $-0.0619\;(0.0134)$ & $-2.32\times10^{-6}\;(3.25\times10^{-6})$ \\
Supplier: outgoing & D & 2 & $0.0121\;(0.0079)$ & $-5.14\times10^{-6}\;(3.11\times10^{-6})$ \\
Supplier: outgoing & D & 3 & $-0.0023\;(0.0082)$ & $-3.83\times10^{-7}\;(1.15\times10^{-6})$ \\
Supplier: outgoing & D & 4 & $-0.0122\;(0.0076)$ & $-4.62\times10^{-7}\;(1.18\times10^{-6})$ \\
Supplier: outgoing & D & 5 & $-0.0118\;(0.0051)$ & $7.98\times10^{-7}\;(8.39\times10^{-7})$ \\
\noalign{\vskip 3pt}
Supplier: outgoing & F & 0 & $0.0144\;(0.0057)$ & $8.79\times10^{-5}\;(1.48\times10^{-5})$ \\
Supplier: outgoing & F & 1 & $-0.0058\;(0.0012)$ & $-6.99\times10^{-8}\;(5.60\times10^{-6})$ \\
Supplier: outgoing & F & 2 & $-5.78\times10^{-4}\;(5.09\times10^{-4})$ & $5.48\times10^{-6}\;(4.74\times10^{-6})$ \\
Supplier: outgoing & F & 3 & $-6.49\times10^{-4}\;(4.37\times10^{-4})$ & $-2.13\times10^{-6}\;(1.05\times10^{-6})$ \\
Supplier: outgoing & F & 4 & $-0.0016\;(4.91\times10^{-4})$ & $5.87\times10^{-7}\;(1.54\times10^{-6})$ \\
Supplier: outgoing & F & 5 & $-8.40\times10^{-4}\;(4.52\times10^{-4})$ & $1.64\times10^{-7}\;(1.42\times10^{-6})$ \\
\bottomrule
\end{tabular}
\end{table}

\begin{table}[p]
\centering\fontsize{10}{12}\selectfont
\caption{Distributed-lag coefficients for neighboring churn at the focal client. Entries are coefficient (standard error), both in percentage points per additional neighboring event. Supplier and client tables partition the same jointly estimated model: each focal outcome includes all 48 neighboring counts, controls and supplier-role, client-role and year effects. Standard errors use supplier/client two-way clustering. Coefficient probabilities are not adjusted for multiple comparisons.}
\label{tab:neighbor-lag-client}
\begin{tabular}{@{}>{\raggedright\arraybackslash}p{3.2cm}ccrr@{}}
\toprule
Position & Neighbor & Lag & Focal D & Focal F \\
\midrule
Client: incoming & D & 0 & $0.0242\;(0.0063)$ & $-2.47\times10^{-6}\;(4.09\times10^{-6})$ \\
Client: incoming & D & 1 & $3.06\times10^{-4}\;(0.0025)$ & $-6.01\times10^{-6}\;(4.48\times10^{-6})$ \\
Client: incoming & D & 2 & $5.93\times10^{-4}\;(0.0017)$ & $1.44\times10^{-6}\;(3.44\times10^{-6})$ \\
Client: incoming & D & 3 & $0.0036\;(0.0017)$ & $-3.88\times10^{-6}\;(3.61\times10^{-6})$ \\
Client: incoming & D & 4 & $0.0045\;(0.0017)$ & $-4.45\times10^{-6}\;(3.01\times10^{-6})$ \\
Client: incoming & D & 5 & $-0.0074\;(0.002)$ & $-2.21\times10^{-6}\;(2.81\times10^{-6})$ \\
\noalign{\vskip 3pt}
Client: incoming & F & 0 & $0.015\;(0.0031)$ & $7.00\times10^{-5}\;(6.45\times10^{-6})$ \\
Client: incoming & F & 1 & $-0.0086\;(0.0017)$ & $1.94\times10^{-6}\;(4.20\times10^{-6})$ \\
Client: incoming & F & 2 & $-0.0028\;(7.59\times10^{-4})$ & $3.31\times10^{-6}\;(3.53\times10^{-6})$ \\
Client: incoming & F & 3 & $-0.0036\;(7.67\times10^{-4})$ & $3.25\times10^{-6}\;(5.46\times10^{-6})$ \\
Client: incoming & F & 4 & $-0.0046\;(6.92\times10^{-4})$ & $-8.49\times10^{-7}\;(3.39\times10^{-6})$ \\
Client: incoming & F & 5 & $-0.0037\;(8.91\times10^{-4})$ & $-3.44\times10^{-7}\;(2.53\times10^{-6})$ \\
\midrule
Client: outgoing & D & 0 & $0.0774\;(0.0232)$ & $7.86\times10^{-6}\;(3.11\times10^{-6})$ \\
Client: outgoing & D & 1 & $-0.0252\;(0.0079)$ & $-1.81\times10^{-6}\;(1.80\times10^{-6})$ \\
Client: outgoing & D & 2 & $0.0137\;(0.0075)$ & $2.35\times10^{-7}\;(1.79\times10^{-6})$ \\
Client: outgoing & D & 3 & $-0.0061\;(0.0054)$ & $1.24\times10^{-6}\;(1.21\times10^{-6})$ \\
Client: outgoing & D & 4 & $-0.0098\;(0.0043)$ & $2.64\times10^{-6}\;(1.32\times10^{-6})$ \\
Client: outgoing & D & 5 & $-0.0017\;(0.0026)$ & $2.10\times10^{-6}\;(1.10\times10^{-6})$ \\
\noalign{\vskip 3pt}
Client: outgoing & F & 0 & $0.0117\;(0.0043)$ & $-3.56\times10^{-6}\;(5.40\times10^{-6})$ \\
Client: outgoing & F & 1 & $-0.0019\;(0.0013)$ & $1.36\times10^{-7}\;(1.55\times10^{-6})$ \\
Client: outgoing & F & 2 & $0.0012\;(7.69\times10^{-4})$ & $1.24\times10^{-7}\;(2.38\times10^{-6})$ \\
Client: outgoing & F & 3 & $6.48\times10^{-4}\;(5.86\times10^{-4})$ & $-1.59\times10^{-7}\;(1.02\times10^{-6})$ \\
Client: outgoing & F & 4 & $0.0019\;(5.08\times10^{-4})$ & $4.70\times10^{-8}\;(1.59\times10^{-6})$ \\
Client: outgoing & F & 5 & $0.0023\;(7.44\times10^{-4})$ & $4.11\times10^{-7}\;(5.72\times10^{-7})$ \\
\bottomrule
\end{tabular}
\end{table}
\FloatBarrier

\FloatBarrier
\section{Joint excess across years}
\suppressfloats[t]
\label{app:joint-multiyear}
We extend the two comparisons in Analysis~3 (joint excess churn; Section~\ref{sec:joint-excess}) by pairing a firm's excess in one year with its transaction neighbor's excess one to five years later. This asks about cross-year alignment between firms; it does not require both firms to exceed together again in the later year.

\subsection{Year separation, backgrounds and eligible observations}
For separation $k\in\{1,\ldots,5\}$, starting years run from 2013 to $2024-k$. We use $\mathcal P_t$, the set of pairs linked by a transaction in $t-1$ or $t$. No transaction is required in the later year. The sensitivity background uses the whole-period pair union $\mathcal U$ from Analysis~3 (joint excess churn; Section~\ref{sec:joint-excess}). In both cases, both firms must have eligible records for the churn intervals ending in $t$ and $t+k$. They need not be eligible in all intervening years or throughout the study period. Missing firm-years are excluded, never assigned zero excess.

Let $\mathcal P^{g}_{t,k}$ denote the retained pairs for background $g$, and let $M_{g,k}=\sum_t|\mathcal P^{g}_{t,k}|$. Because a background pair is unordered, we average the two temporal orders. Using the existing exceedance indicators and positive excesses in Equations~\eqref{eq:exceedance-indicator} and~\eqref{eq:excess}, the occurrence frequency is
\begin{equation}
F_{g,k}=\frac{1}{2M_{g,k}}\sum_t\sum_{\{u,v\}\in\mathcal P^{g}_{t,k}}
\left(Z_{ut}Z_{v,t+k}+Z_{vt}Z_{u,t+k}\right),
\label{eq:joint-lag-frequency}
\end{equation}
and overall amount is
\begin{equation}
T_{g,k}=\frac{1}{2M_{g,k}}\sum_t\sum_{\{u,v\}\in\mathcal P^{g}_{t,k}}
\left[\min(e_{ut},e_{v,t+k})+\min(e_{vt},e_{u,t+k})\right].
\label{eq:joint-lag-amount}
\end{equation}
Zero outcomes remain in both denominators. Each term pairs an outcome from one firm with an outcome from its neighbor.

The eligible years and pairs change with $k$. For comparison, we also evaluate a matched same-year statistic on exactly $\mathcal P^{g}_{t,k}$, replacing each later-year variable in Equations~\eqref{eq:joint-lag-frequency}--\eqref{eq:joint-lag-amount} with its start-year value. These matched ratios describe contemporaneous alignment in the same starting sample; they do not use pairs newly selected from the later-year network.

\subsection{Reference comparisons and tests}
To obtain the expected frequency under the binary reference, we replace the two products in Equation~\eqref{eq:joint-lag-frequency} with $q_{ut}q_{v,t+k}$ and $q_{vt}q_{u,t+k}$. For the occurrence tests, we use the 1,999 simulations of firm-year exceedance indicators from Analysis~3. The amount tests use the 1,999 count allocations from the same analysis, which hold each firm's total fixed across its eligible years.

Within each simulation, we use the same value for a given firm-year wherever it enters a comparison, including pairs sharing a firm and comparisons at different separations. At separation zero, both statistics give the same-year comparison for the corresponding background.

For each positive separation, the upper-tail Monte Carlo probability is one plus the number of simulated statistics at least as large as observed, divided by 2,000. Holm correction covers ten occurrence comparisons, comprising five separations and two backgrounds, and separately ten amount comparisons. Both corrections cover only positive separations.

\subsection{Results}
On the start-year transaction background, the occurrence ratio decreases from 1.089 at a one-year separation to 1.036 at five years, while the overall-amount ratio decreases from 1.332 to 1.244 (Tables~\ref{tab:joint-lag-occurrence} and~\ref{tab:joint-lag-amount}). The eligible pair-start-year count decreases from 51,845,416 to 28,362,720. Every positive-lag comparison for both measures has upper-tail Monte Carlo $p=0.0005$, the minimum attainable value, and Holm-adjusted $p=0.005$ in its corresponding ten-comparison family.

\begin{figure}[p]
\centering
\includegraphics[width=\linewidth,height=0.80\textheight,keepaspectratio]{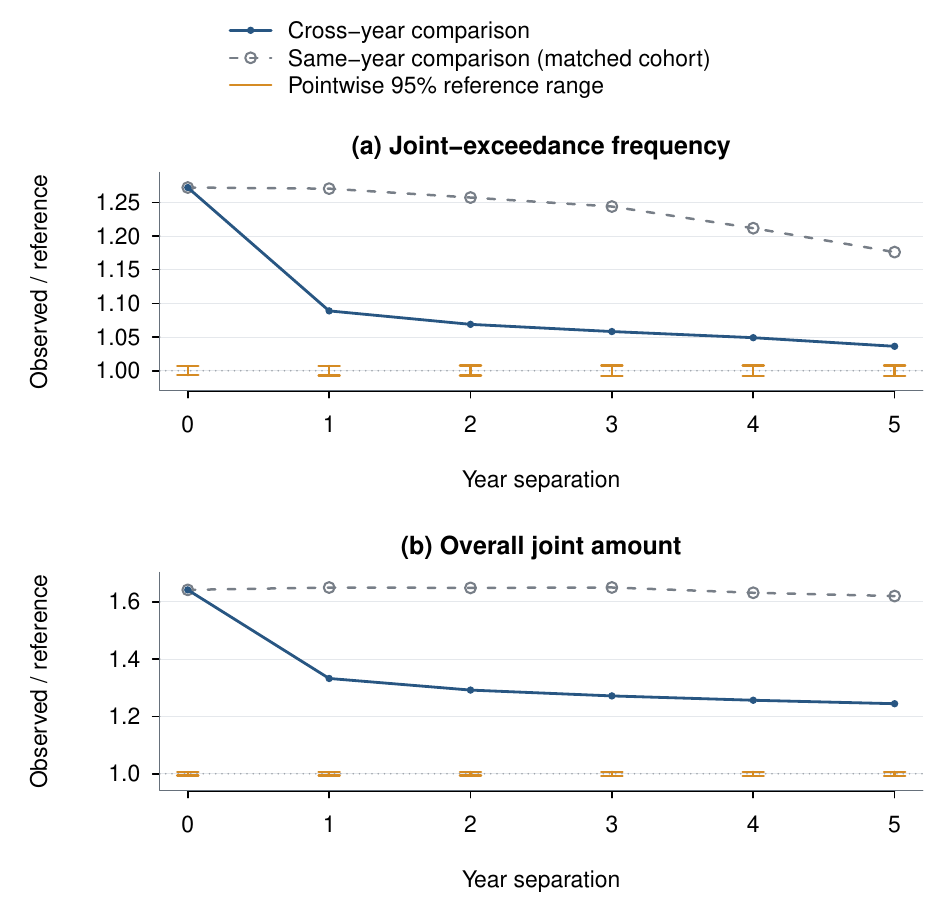}
\caption{Joint excess across years on the start-year transaction background. The across-year ratio compares a firm's year-$t$ outcome with its neighbor's year-$t+k$ outcome, averaging both temporal orders. The matched same-year ratio uses the identical eligible starting pairs at each separation. Occurrence uses the analytic binary expectation, and overall amount uses the mean of 1,999 fixed-total count allocations. Whiskers show pointwise central 95\% ranges of the simulated statistics divided by their reference means. Tables~\ref{tab:joint-lag-occurrence} and~\ref{tab:joint-lag-amount} report the tests at positive lags.}
\label{fig:joint-lag}
\end{figure}

Matched same-year ratios are larger than the across-year ratios at every separation: they range from 1.176 to 1.270 for occurrence and from 1.620 to 1.650 for amount (Figure~\ref{fig:joint-lag}). Thus the stronger contemporaneous alignment is also present when the starting sample is held identical within each comparison. These matched summaries are descriptive; no additional test of a difference between separations is reported.

The broader whole-period-union background gives occurrence ratios of 1.025 and 1.010 at one- and two-year separations, with Holm-adjusted $p=0.005$ and $p=0.012$, respectively. At three to five years the ratios range from 0.978 to 1.001, and none of the pooled occurrence frequencies significantly exceeds its reference. Overall-amount ratios remain above one at all five separations, ranging from 1.174 to 1.255, with Holm-adjusted $p=0.005$ throughout. The background definition therefore matters for the persistence of excess frequency, while excess overall amount remains under both background definitions. These associations do not establish that earlier churn causes later churn.

\begin{table}[p]
\centering\small
\caption{Frequency with which both firms exceed their respective fitted means at year separations zero to five. The two temporal orders are pooled, so observed and expected frequencies divide by twice the eligible pair-start-year count. The matched ratio uses same-year outcomes at the start of exactly the same eligible pair set. Monte Carlo probabilities use 1,999 simulations of the exceedance indicators. Holm correction covers ten positive-lag comparisons across the two backgrounds.}
\label{tab:joint-lag-occurrence}
\begin{tabular}{@{}rrrrrrrr@{}}
\toprule
Lag & Pair-start-years & \shortstack{Observed\\(\%)} & \shortstack{Expected\\(\%)} & \shortstack{Across-year\\ratio} & \shortstack{Matched\\same-year ratio} & MC $p$ & Holm $p$ \\
\midrule
\multicolumn{8}{@{}l@{}}{\textit{Start-year transaction neighbors}} \\[3pt]
0 & 59,099,713 & 15.773 & 12.400 & 1.272 & 1.272 & -- & -- \\
1 & 51,845,416 & 13.565 & 12.458 & 1.089 & 1.270 & 0.0005 & 0.005 \\
2 & 45,171,057 & 13.389 & 12.526 & 1.069 & 1.257 & 0.0005 & 0.005 \\
3 & 39,093,509 & 13.280 & 12.550 & 1.058 & 1.244 & 0.0005 & 0.005 \\
4 & 33,529,298 & 13.247 & 12.627 & 1.049 & 1.212 & 0.0005 & 0.005 \\
5 & 28,362,720 & 13.259 & 12.795 & 1.036 & 1.176 & 0.0005 & 0.005 \\
\midrule
\multicolumn{8}{@{}l@{}}{\textit{Whole-period pair union}} \\[3pt]
0 & 84,057,287 & 14.576 & 13.239 & 1.101 & 1.101 & -- & -- \\
1 & 74,433,329 & 13.597 & 13.265 & 1.025 & 1.097 & 0.0005 & 0.005 \\
2 & 65,366,534 & 13.443 & 13.309 & 1.010 & 1.084 & 0.003 & 0.012 \\
3 & 57,117,226 & 13.327 & 13.312 & 1.001 & 1.069 & 0.370 & 1.000 \\
4 & 49,476,439 & 13.260 & 13.367 & 0.992 & 1.038 & 0.987 & 1.000 \\
5 & 42,253,694 & 13.215 & 13.517 & 0.978 & 1.002 & 1.000 & 1.000 \\
\bottomrule
\end{tabular}
\end{table}

\begin{table}[p]
\centering\small
\caption{Overall joint excess amount at year separations zero to five, retaining zero outcomes. The reference is the mean of 1,999 fixed-total count allocations. The matched ratio evaluates same-year excesses at the start of the identical eligible pair set. Holm correction covers ten positive-lag amount comparisons across two backgrounds.}
\label{tab:joint-lag-amount}
\begin{tabular}{@{}rrrrrrrr@{}}
\toprule
Lag & Pair-start-years & \shortstack{Observed\\mean} & \shortstack{Reference\\mean} & \shortstack{Across-year\\ratio} & \shortstack{Matched\\same-year ratio} & MC $p$ & Holm $p$ \\
\midrule
\multicolumn{8}{@{}l@{}}{\textit{Start-year transaction neighbors}} \\[3pt]
0 & 59,099,713 & 0.23909 & 0.14566 & 1.641 & 1.641 & -- & -- \\
1 & 51,845,416 & 0.19710 & 0.14793 & 1.332 & 1.650 & 0.0005 & 0.005 \\
2 & 45,171,057 & 0.19404 & 0.15021 & 1.292 & 1.648 & 0.0005 & 0.005 \\
3 & 39,093,509 & 0.19293 & 0.15173 & 1.272 & 1.650 & 0.0005 & 0.005 \\
4 & 33,529,298 & 0.19234 & 0.15309 & 1.256 & 1.631 & 0.0005 & 0.005 \\
5 & 28,362,720 & 0.19278 & 0.15493 & 1.244 & 1.620 & 0.0005 & 0.005 \\
\midrule
\multicolumn{8}{@{}l@{}}{\textit{Whole-period pair union}} \\[3pt]
0 & 84,057,287 & 0.21747 & 0.15760 & 1.380 & 1.380 & -- & -- \\
1 & 74,433,329 & 0.20015 & 0.15949 & 1.255 & 1.384 & 0.0005 & 0.005 \\
2 & 65,366,534 & 0.19751 & 0.16148 & 1.223 & 1.380 & 0.0005 & 0.005 \\
3 & 57,117,226 & 0.19608 & 0.16272 & 1.205 & 1.376 & 0.0005 & 0.005 \\
4 & 49,476,439 & 0.19473 & 0.16381 & 1.189 & 1.354 & 0.0005 & 0.005 \\
5 & 42,253,694 & 0.19415 & 0.16542 & 1.174 & 1.336 & 0.0005 & 0.005 \\
\bottomrule
\end{tabular}
\end{table}

\FloatBarrier
\end{document}